%% file: main.tex
\pdfoutput=1
\documentclass[english]{article}
\usepackage{geometry}
\usepackage{babel}
\usepackage{xcolor}
\usepackage{graphicx}
\usepackage{amsmath,amsfonts}
\usepackage{xcolor}
\usepackage{mathcomp}
\usepackage{textcomp}
\usepackage{amssymb}
\usepackage{graphicx,indentfirst,subfigure,epsfig}
 \usepackage{varioref}
 \usepackage{wrapfig}
 \usepackage{subfigure}
 \usepackage{subfigmat}
 \usepackage{amsthm}
\input{pauls-macros}

\usepackage{hyperref}
\hypersetup{colorlinks, citecolor=red, linkcolor=blue, urlcolor=red, breaklinks=true}
\usepackage{scrextend}
\usepackage[T1]{fontenc}
\usepackage[utf8]{inputenc}
\usepackage{tgtermes}
\usepackage{nomencl}
\usepackage{algorithm}
\usepackage{algorithmic}
\usepackage{booktabs}
\usepackage{longtable}
\usepackage{lineno}
\makenomenclature

\begin{document}
\large
\title{\textbf{Supporting Multi-point Fan Design with Dimension Reduction}}
\author{Pranay Seshadri\footnote{Address all correspondence to \texttt{ps583@cam.ac.uk}}$\;^{\; \star}$, Shaowu Yuchi$^{\star}$, Shahrokh Shahpar$^{\dagger}$,  Geoffrey Parks$^{\star}$\\ \vspace{0.3 cm} \\ $^{\star}$Department of Engineering, University of Cambridge, Cambridge, U. K., \\$^{\dagger}$Central Technology, Rolls-Royce plc., Derby, U. K.}
\date{}
\maketitle{}
\begin{abstract}
Motivated by the idea of turbomachinery active subspace performance maps, this paper studies dimension reduction in turbomachinery 3D CFD simulations. First, we show that these subspaces exist across different blades---under the same parametrization---largely independent of their Mach number or Reynolds number. This is demonstrated via a numerical study on three different blades. Then, in an attempt to reduce the computational cost of identifying a suitable dimension reducing subspace, we examine statistical sufficient dimension reduction methods, including sliced inverse regression, sliced average variance estimation, principal Hessian directions and contour regression. Unsatisfied by these results, we evaluate a new idea based on polynomial variable projection---a non-linear least squares problem. Our results using polynomial variable projection clearly demonstrate that one can accurately identify dimension reducing subspaces for turbomachinery functionals at a fraction of the cost associated with prior methods. We apply these subspaces to the problem of comparing design configurations across different flight points on a working line of a fan blade. We demonstrate how designs that offer a healthy compromise between performance at cruise and sea-level conditions can be easily found by visually inspecting their subspaces.
\end{abstract}

\printnomenclature
\input{sec-1}
\input{sec-2}
\input{sec-3}

\input{sec-5}

\input{sec-6}

\input{sec-7}
\bibliographystyle{abbrv}
\bibliography{references}

\end{document}

%% file: pauls-macros.tex
\usepackage{bm}

  \def\clap#1{\hbox to 0pt{\hss#1\hss}}

\providecommand{\mat}[1]{\bm{#1}}%
\renewcommand{\vec}[1]{\mathbf{#1}}

\providecommand{\mA}{\ensuremath{\mat{A}}}

\providecommand{\mI}{\ensuremath{\mat{I}}}
\providecommand{\mJ}{\ensuremath{\mat{J}}}
\providecommand{\mK}{\ensuremath{\mat{K}}}

\providecommand{\mP}{\ensuremath{\mat{P}}}

\providecommand{\mU}{\ensuremath{\mat{U}}}
\providecommand{\mV}{\ensuremath{\mat{V}}}
\providecommand{\mW}{\ensuremath{\mat{W}}}

\providecommand{\vc}{\ensuremath{\vec{c}}}

\providecommand{\vf}{\ensuremath{\vec{f}}}

\providecommand{\vh}{\ensuremath{\vec{h}}}

\providecommand{\vr}{\ensuremath{\vec{r}}}

\providecommand{\vu}{\ensuremath{\vec{u}}}
\providecommand{\vv}{\ensuremath{\vec{v}}}
\providecommand{\vw}{\ensuremath{\vec{w}}}
\providecommand{\vx}{\ensuremath{\vec{x}}}
\providecommand{\vy}{\ensuremath{\vec{y}}}
\providecommand{\vz}{\ensuremath{\vec{z}}}



%% file: sec-1.tex
\section{Introduction}
In the modern turbomachinery design process, blade shapes are parametrized by a set of design variables useful from aerodynamic, structural and manufacturing perspectives. Typically, the number of such parameters can range from 20 to 250. This high-dimensional design space makes parametric studies---e.g., uncertainty quantification, optimization and sensitivity analysis---challenging and computationally prohibitive, motivating the use of dimension reduction strategies. Within the broad field of dimension reduction there exist two basic approaches: subspace- and subset-based dimension reduction. Subset-based dimension reduction aims to find an important\footnote{Importance here is usually characterized by the conditional variance on the output, although other metrics can also be adopted.} subset of original variables. Methods such as ridge regression, Lasso \cite{friedman2001elements} and ANOVA-based decompositions and the related Sobol' indices \cite{sobol2001global, archer1997sensitivity} can be categorized as methods for subset-based dimension reduction. Our interest in this paper, however, lies in ideas for subspace-based dimension reduction, where one seeks to identify specific linear combinations of all the parameters that are important with respect to the output. 

This body of work by no means represents the first foray into turbomachinery-based dimension reduction. Prior papers that investigated the use of dimension reduction strategies in turbomachinery design include the works of Banamonde et al.~\cite{bahamonde2017active} and Qin et al.~\cite{qin2016flow}. In the former publication, the authors use \emph{active subspaces} \cite{constantine2015active}---a set of ideas for parameter space dimension reduction---to develop a reduced-order model for the design of a turbine power cycle. To compute the active subspaces, the authors used finite differences, a feasible approach given that their governing model was a mean-line code. In \cite{qin2016flow} the authors use a 2D CFD model of a NACA65-1210 airfoil and leverage a one-dimensional active subspaces heuristic (see page 5 in \cite{constantine2015active}) to estimate the impact of blade fouling and erosion. 

More recently (and of greater relevance to this work), Seshadri et al.~\cite{seshadri2018turbomachinery} detail methods for generating \emph{turbomachinery active subspace performance maps}. These 2D contour plots illustrate and quantify the variation of key flow performance metrics with different designs---even when the number of design variables is greater than two. The authors generate such maps for an isolated 3D fan blade (see Fig.~24 in \cite{seshadri2018turbomachinery}), where the contours reflect the changes in pressure ratio, flow capacity and efficiency with a change in the design of the blade. Their work raises several critical questions---both from a computational perspective and from a turbomachinery design standpoint----motivating our paper. As the study in \cite{seshadri2018turbomachinery} was carried out on a single isolated blade, it is logical to ask whether those results can be generalized to other blades---even assuming the same design parametrization. We explore this line of questioning in Sect.~\ref{sec:one} on three different modern fan blades. The computational workhorse for computing the dimension reducing subspaces in \cite{seshadri2018turbomachinery} was an active subspace technique that leverages a global quadratic polynomial approximation. In Sect.~\ref{sec:two} we explore a variety of alternative classical approaches. Motivated by the variable projection aided dimension reduction approach in \cite{hokanson2017data}, we devote Sect.~\ref{sec:three} to its study, offering a few minor algorithmic variations. Then, we test this on one of our blades, being parsimonious in the number of samples (and hence CFD evaluations) used. As turbomachinery components need to operate at multiple points across a characteristic, in Sect.~\ref{sec:four} we leverage variable projection and a pressure ratio interpolating strategy to obtain design maps at two points on the characteristic and discuss their utility within the design process.

%% file: sec-2.tex
\section{Turbomachinery Active Subspace Performance Maps}
\label{sec:one}
\begin{figure}
\centering
\includegraphics[scale=0.5]{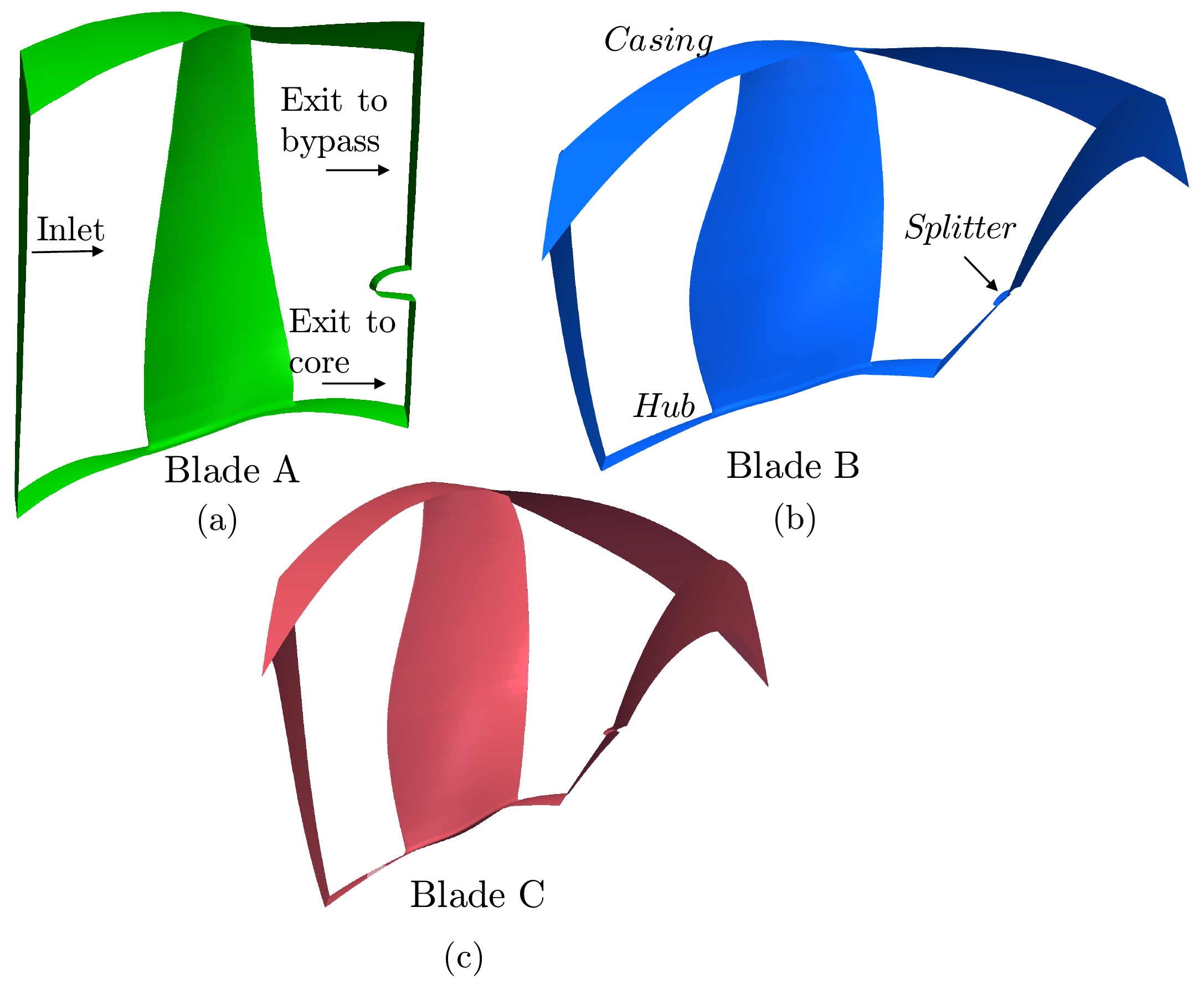}
\caption{The three fan blades considered in this study (images have been appropriately scaled for proprietary reasons).}
\label{fanblades}
\end{figure}

\begin{figure}
\centering
\includegraphics[scale=0.4]{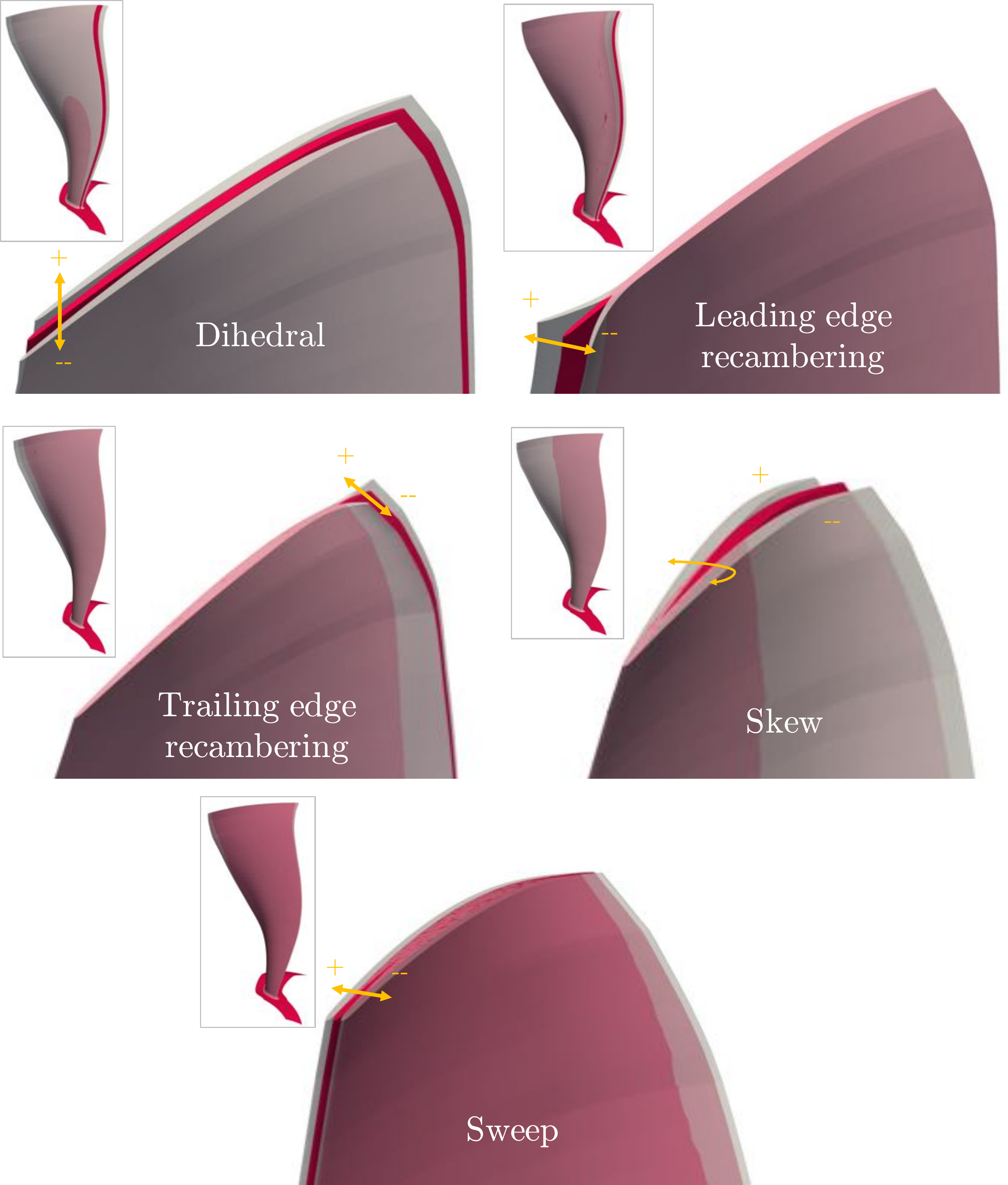}
\caption{The design space of Blades A, B and C. Shown are the five degrees of freedom, where the ranges have been exaggerated for clarity.}
\label{design_space}
\end{figure}

In this study, three research fan blades are considered, shown in Fig.~\ref{fanblades}. Blades A and B are high-speed fan blades, with real engine representative pressure ratios and Mach numbers, while Blade C is a low-speed research fan blade.
Each blade uses the same 25D parametrization scheme in \cite{seshadri2018turbomachinery}, defined by 5 degrees of freedom (dof) specified at 5 spanwise locations. These dofs include leading and trailing edge recambering, skew, sweep and dihedral as shown in Fig.~\ref{design_space}.

\subsection{Computational grid and boundary conditions}
\begin{figure}[h]
\centering
\includegraphics[scale=0.5]{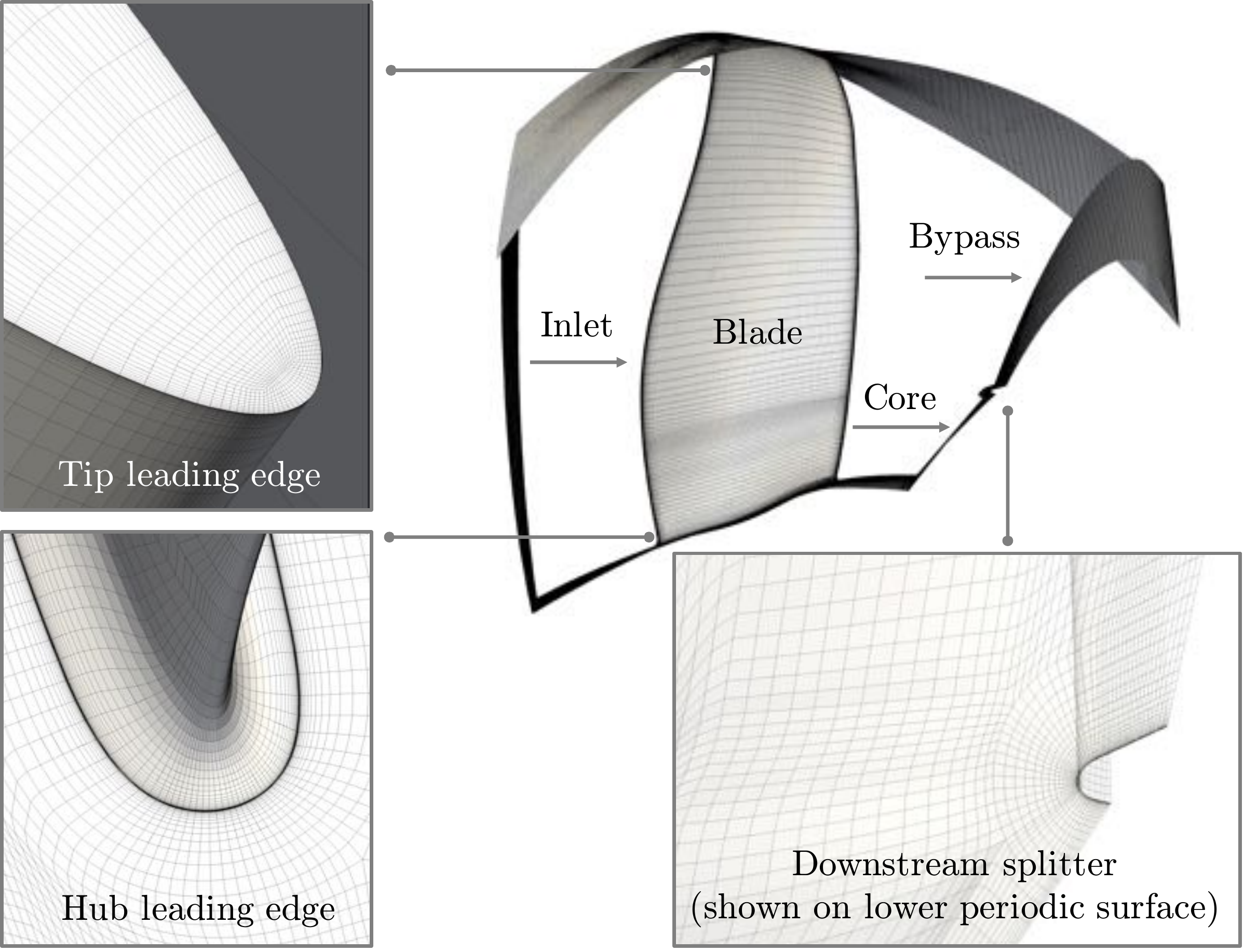}
\caption{Representative mesh topology used for all the blades (shown here for Blade C).}
\label{mesh_padram}
\end{figure}

The Rolls-Royce code PADRAM \cite{PADRAM} was used for both generating the geometry and for creating its corresponding computational grid, as shown in Fig.~\ref{mesh_padram}. The code is an algebraic-elliptic, multi-block grid generator based on a C-O-H topology. For all three fan blades, the mesh consisted of downstream and upstream H blocks for the stationary zones, upper and lower H blocks for the periodic boundaries, an O-mesh around the blade, and a C-mesh for the downstream splitter. The meshes were created with eleven blocks each; details of the meshes for the three blades is provided in Table \ref{table:mesh_characteristics}. 

\begin{table}
\begin{center}
\caption{Mesh characteristics for the three blades.}
\label{table:mesh_characteristics}
\begin{tabular}{llll}
\hline
  \textbf{Property} & \textbf{Blade A} & \textbf{Blade B} & \textbf{Blade C} \\
\hline
  No. of cells in the tip gap & 14 & 10 & 14\\
  No. of cells in the spanwise direction & 129 & 159 & 139 \\
  No. of cells in the circumferential direction & 59 & 55 & 54 \\
  No. of cells in the axial direction & 177 & 323 & 281 \\
  Total number of cells (millions) & 1.7 & 3.345 & 3.153\\ \hline
\end{tabular}
\end{center}
\end{table}

\begin{figure}
\begin{subfigmatrix}{2}
\subfigure[]{\includegraphics{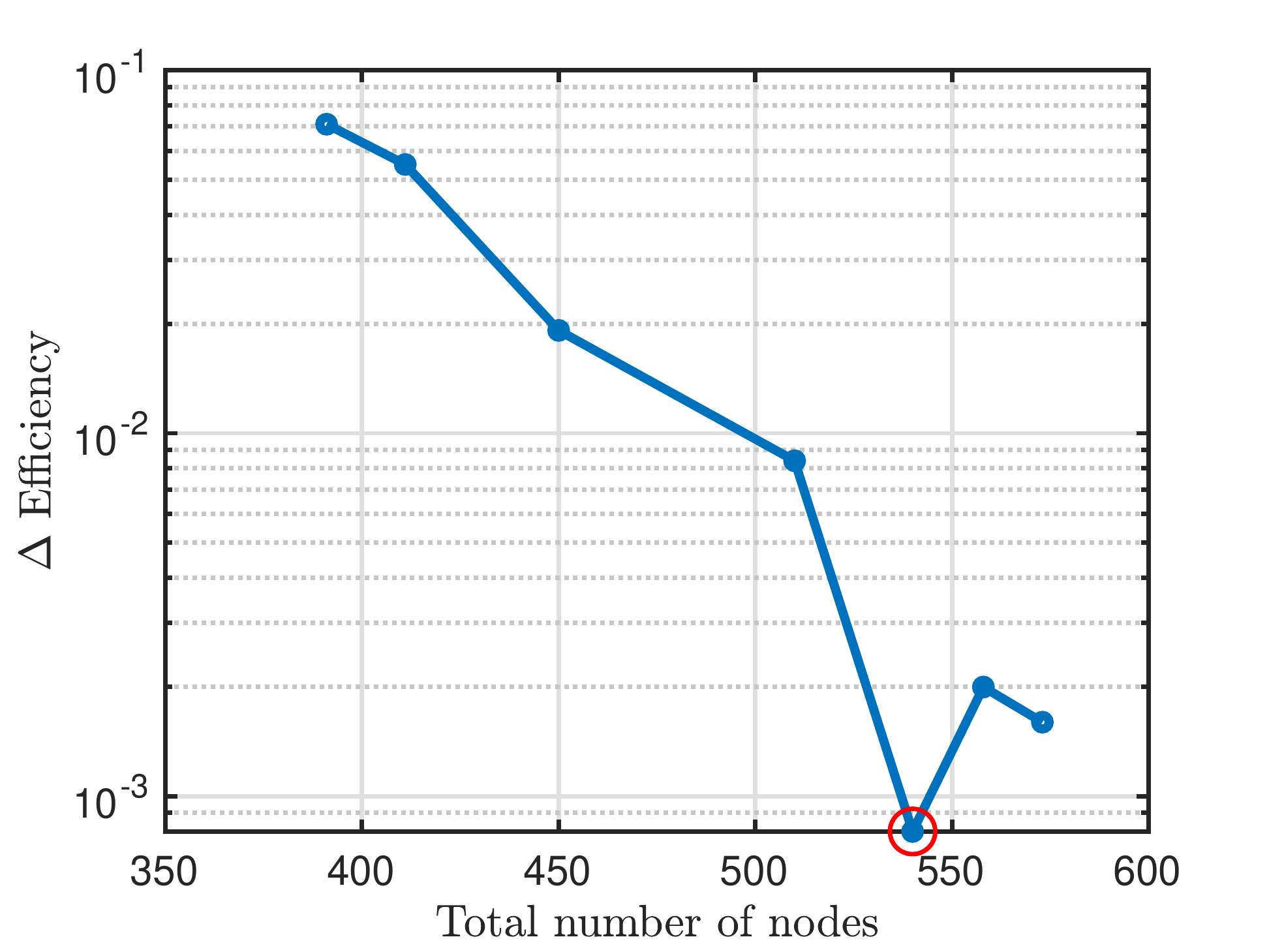}}
\subfigure[]{\includegraphics{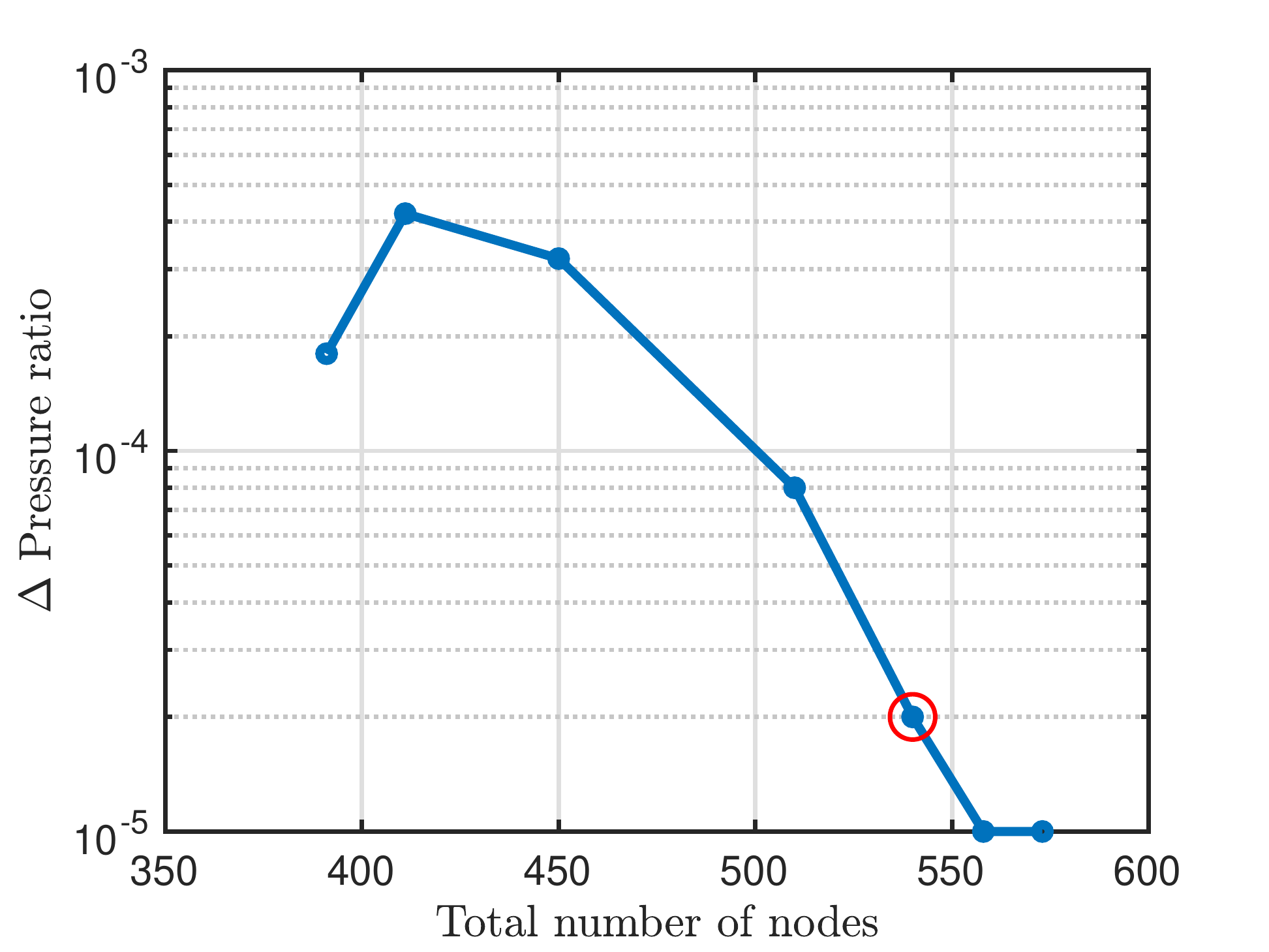}}
\end{subfigmatrix}
\caption{Mesh convergence plots for Blade C at cruise. Shown are the objectives of (a) non-dimensional efficiency and (b) non-dimensional pressure ratio. The mesh with the red circular maker was used in our computations.}
\label{mesh_convergence}
\end{figure}

A grid independence study was undertaken on all three meshes. For completeness, we provide details of the mesh convergence study undertaken on Blade C at cruise. The number of cells in the spanwise, circumferential and axial direction were varied for convergence in the pressure ratio and efficiency, see Figure~\ref{mesh_convergence}. 

These are defined by
\begin{equation}
PR=\left(\frac{\dot{m}_{\text{byp}} P_{\text{byp}}+ \dot{m}_{\text{core}}P_{\text{core}}}{ \dot{m}_{\text{in}}P_{\text{in}}}\right)
\label{pr_eq}
\end{equation}
and
\begin{equation}
\eta=\frac{\left(PR^{\left(\gamma-1\right)/\gamma}-1\right)}{TR-1}\times100
\label{eff_eq}
\end{equation}
where the temperature ratio ($TR$) is given by
\begin{equation}
TR=\left(\frac{\dot{m}_{\text{byp}}T_{\text{byp}}+ \dot{m}_{\text{core}}T_{\text{core}}}{\dot{m}_{\text{in}}T_{\text{in}}}\right)
\label{tr}
\end{equation}
In Eqs.~\eqref{pr_eq}, \eqref{eff_eq} and \eqref{tr}: $P$ indicates the stagnation pressure; $T$ the stagnation temperature; $\dot{m}$ the mass-flow rate, and the subscripts `byp', `core' and `in' indicate flow quantities at the bypass outlet, core outlet and inlet respectively. We use mass-averaged values (radially and circumferentially) of the stagnation temperature and pressure obtained at the three planes indicated above. In terms of boundary conditions, at the inlet we enforce a fixed stagnation pressure, and at the exit to both the core and the bypass we utilize a fixed exit capacity $\Gamma$ boundary condition:
\begin{equation}
\Gamma_{\text{core}}=\dot{m}_{\text{core}}\frac{\sqrt{T_{\text{core}}}}{P_{\text{core}}}
\end{equation}
\begin{equation}
\Gamma_{\text{byp}}=\dot{m}_{\text{byp}}\frac{\sqrt{T_{\text{byp}}}}{P_{\text{byp}}}.
\end{equation}
The endwalls (hub, blade, casing and splitter) are characterized by viscous wall boundary conditions with both the hub and the blade having an angular velocity based on the shaft speed. 

The mesh convergence study involved 8 different meshes, each with different number of grid points. The number of points in the grid were varied by increasing (i) axial nodes; (ii) circumferential nodes; (iii) spanwise nodes (including the tip gap) and their distribution. Careful attention was also paid to the nodes where the O-mesh of the blade mated with the upper and lower H-meshes. The results in Figure~\ref{mesh_convergence}(a) plot the $\Delta$ Efficiency on the horizontal axis (logarithm-base 10 scale), defined as
\begin{equation}
\Delta \; \text{Efficiency} = \left| \text{Efficiency}_{\text{finest grid} } - \text{Efficiency}_{\text{current grid} } \right|.
\end{equation}
Similarly, in (b), $\Delta$ Pressure ratio is given by
\begin{equation}
\Delta \; \text{Pressure ratio} = \left| \text{Pressure ratio}_{\text{finest grid} } - \text{Pressure ratio}_{\text{current grid} } \right| .  
\end{equation}
The red circles in these plots indicate the mesh used in all our subsequent computations. For this mesh, we have a confidence in the first two decimal places in efficiency and the first four decimal places in pressure ratio.
\subsection{Computational flow solver and constitutive equations}
We solve the 3D Reynolds-averaged Navier-Stokes (RANS) equations on the aforementioned PADRAM meshes, using the HYDRA flow solver \cite{crumpton1998unstructured, lapworth2004hydra}. To close the set of equations when applying a Reynolds-based averaging to the Navier-Stokes, we utilize the one equation Spalart-Allmaras turbulence model \cite{spalart1992one}. Further details on the calculation of the inviscid fluxes, the viscous fluxes and convergence can be found in pages 10-21 of \cite{ostling2011potential}. We omit a detailed flow validation here and direct interested readers to references \cite{Giulio, Seshadri_LEAK}, where experiment vs.~CFD validations for the PADRAM-HYDRA suite of tools are provided for representative transonic fan blades. Boundary conditions and the inlet Reynolds number for the four design of experiments carried out in this study are given in Table~\ref{table:Table_BCs}.

\begin{table}
\begin{center}
\caption{Reynolds number and inlet boundary conditions.}
\label{table:Table_BCs}
\begin{tabular}{lllll}
\hline
  \textbf{Property} & \textbf{Blade A} & \textbf{Blade B} & \textbf{Blade C (cruise) } & \textbf{Blade C (sea-level) } \\
\hline
$P_{in} \;  (N/m^2) $ &   $1.005 \times 10^{5}$  &      $1.011 \times 10^{5}$  & $3.14 \times 10^{4}$ &   $1.005 \times 10^{5}$  \\
$T_{in} \; (K)$ &     $288$   &     $288$     &   $248$    &  $288$ \\
$Re$ &  $7.8 \times 10^{6} $ & $1.8 \times 10^{6} $   & $3.41 \times 10^{6} $ &  $8.47 \times 10^{6} $  \\ \hline
\end{tabular}
\end{center}
\end{table}

\subsection{Computing active subspaces via a quadratic model}
\begin{figure}
\centering
\begin{subfigmatrix}{2}
\subfigure[]{\includegraphics[]{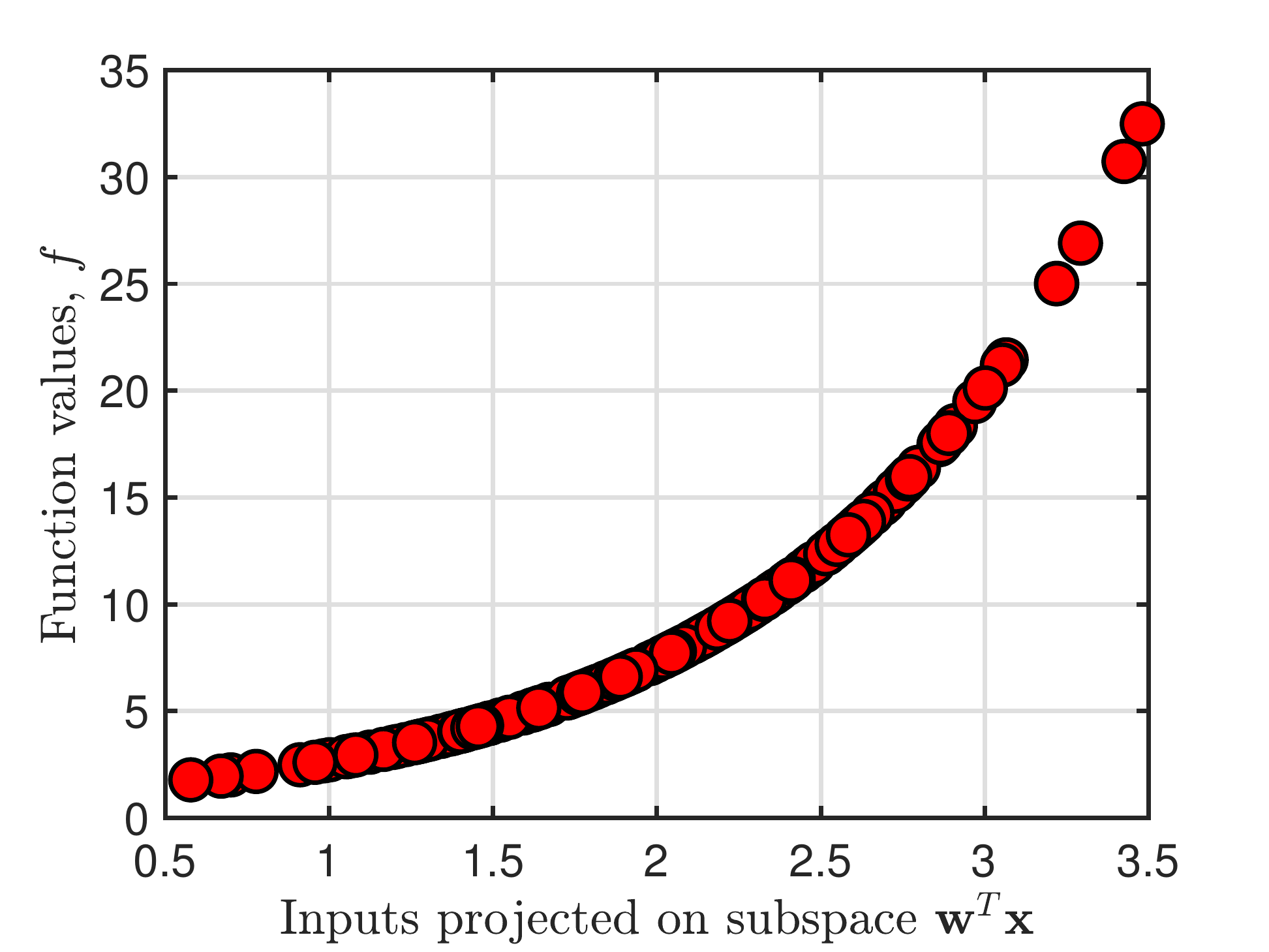}}
\subfigure[]{\includegraphics[]{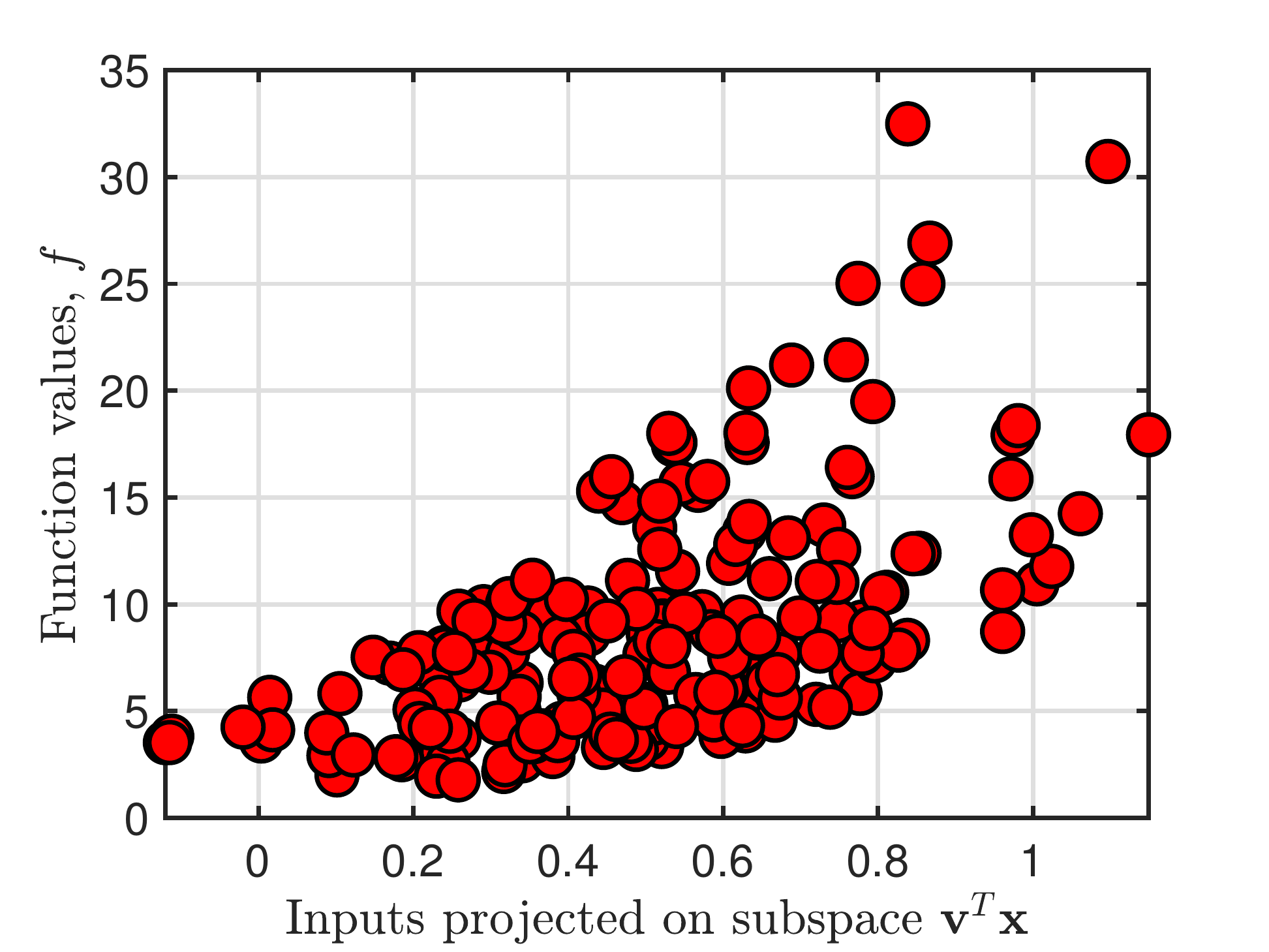}}
\end{subfigmatrix}
\caption{Projecting multi-dimensional samples of the function $\text{exp} \left(x_1 + x_2 + x_3 + x_4 \right)$ over a subspace: (a) $\vw = [1,1,1,1]^{T}$ (b) $\vv = [0.4, -0.3, 0.5, 0.4]^{T}$}
\label{sufficient_summary_plots}
\end{figure}

Before we delve into \emph{active subspaces}, it will be useful to introduce the concept of a \emph{sufficient summary plot} (following the terminology in \cite{cook2009introduction}). Consider the 4D function
\begin{equation}
f \left( \vx \right) = \text{exp} \left(x_1 + x_2 + x_3 + x_4 \right),
\label{equ_exp_func}
\end{equation}
defined over the hypercube $\mathcal{X} \subset [-1, 1]^{4}$ and where $\vx = [x_1, x_2, x_3, x_4]^{T} \in \mathcal{X}$. We can express this function as 
\begin{equation}
f \left( \vx \right) = \text{exp} \left( \vw^{T} \vx \right), \; \; \; \text{where} \; \; \; \vw = [1, 1, 1, 1]^{T}.
\end{equation}
The transformation induced by the vector $\vw$ projects the 4D values in $\vx$ onto the \emph{subspace}---i.e., linear combination of the parameters---given by $\vw^{T} \vx$. Thus, our 4D function is actually 1D over this subspace! Imagine we generate $N$ random samples from $\mathcal{X}$ and project each $\vx_{i}$ for $i=1, \ldots, N$ onto this subspace. Then, we evaluate \eqref{equ_exp_func} for each $\vx_{i}$ and plot the resulting data $f_i = f \left( \vx_{i} \right)$. We would arrive at Fig.~\ref{sufficient_summary_plots}(a), where the horizontal axis of plots the subspace $\vw^{T} \vx_{i}$ and the vertical axis plots the function values $f_i$.

Now, typically, we may not know $\vw$ or even have access to the functional form of a model. Thus, we have to use techniques based on input-output pairs of the model to approximate this subspace. But what exactly are we looking for? Contrast Fig.~\ref{sufficient_summary_plots}(a) with (b) where the subspace used for projection is $\vv = [0.4, -0.3, 0.5, 0.4]^{T}$. It is readily apparent that although this subspace captures the monotonic trend of the function, any further inference is far more challenging compared to (a), owing to the scatter in the plot.

The above discussion raises the question of how one can discover such \emph{dimension reducing} subspaces such that \emph{if} the function admits low dimensional structure of the form of \eqref{equ_exp_func}, then we can construct its sufficient summary plots. In what follows, we will assume that our functional output comes from CFD and that it does admit such low dimensional structure.

We now consider $\vx$ as a set of design parameters and assume $\vx \in \mathbb{R}^{25}$. Without loss of generality, we will assume that $\vx$ has been appropriately non-dimensionalized such that $\boldsymbol{-1} \leq \vx \leq \boldsymbol{1}$. Our goal in this subsection is to detail a strategy for reducing the dimensionality of our problem, via
\begin{equation}
f(\vx) \approx g( \mW_{1}^T \vx)
\label{equ_active}
\end{equation}
where $\mW_{1} \in \mathbb{R}^{d \times m}$ where $d$ is the number of design parameters (here 25) and $m << d$ is the reduced dimension. In Eq.~\eqref{equ_active}, $g$ is a suitably chosen $m$-dimensional function. Ideally, one would like $m$ to be either 1 or 2, facilitating easy visualization. However, this may not necessarily be possible and depends on the relationship between $\vx$ and $f$. So, how exactly do we find $\mW_{1}$? Constantine \cite{constantine2015active} lays the theoretical foundations for approximating $\mW_{1}$ and offers numerous practical computational heuristics. In essence, one seeks to estimate the covariance of the gradient, given by
\begin{equation}
\mK=\int \nabla_{\vx} f(\vx)\nabla_{\vx} f(\vx)^{T} \rho (\vx) d \vx
\label{equ_as}
\end{equation}
where $\mK \in \mathbb{R}^{d \times d}$. Here the measure $\rho(\vx)$ is the density of the inputs $\vx$ over their parameter space $\mathcal{X}$. As we can control our sampling strategy, we set $\rho$ to be the uniform measure and generate samples $\vx_{i}$ randomly with respect to the uniform distribution over $\mathcal{X} \in [\boldsymbol{-1}, \boldsymbol{1}]$. Then, one can approximate $\mW_{1}$ using the first few eigenvectors associated with $\mK$
\begin{equation}
\mK=\mW \boldsymbol{ \Lambda} \mW^{T}=\left[\begin{array}{cc}
\mW_{1} & \mW_{2}\end{array}\right]\left[\begin{array}{cc}
\boldsymbol{ \Lambda_{1} }\\
 & \boldsymbol{ \Lambda_{2}}
\end{array}\right]\left[\begin{array}{cc}
\mW_{1} & \mW_{2}\end{array}\right]^{T}
\end{equation}
Intuitively, one can interpret $\mW_{1}$ as the directions along which $\nabla_{\vx} f$ varies the most, and consequently $\mW_{2}$ as the directions along which $\nabla_{\vx} f$ is (close to) zero. 

In the absence of gradients or adjoints (as in \cite{seshadri2018turbomachinery}) one can approximate $\mW_{1}$ by constructing a global quadratic model in $\mathbb{R}^{m}$
\begin{equation}
f\left(\vx_{i}\right) \approx \frac{1}{2} \vx^{T} \mA \vx + \vc^{T} \vx+e
\label{equ:quadratic}
\end{equation}
where the coefficients $\mA \in \mathbb{R}^{d \times d}$, $\vc \in \mathbb{R}^{d}$ and $e \in \mathbb{R}$ are estimated via least squares. The number of coefficients $M$ associated with this quadratic model is given by
\begin{equation}
N = \left(\begin{array}{c}
d+k\\
k
\end{array}\right),
\label{equ:total_order_scaling}
\end{equation}
where $k$ is the highest degree of the polynomial and $d$ once again corresponds to the dimension of the space where the polynomial has to be constructed. As we seek to \emph{approximate} rather than \emph{interpolate}, the number of CFD evaluations $N$ required to fit our model should be 
\begin{equation}
N = \alpha M
\end{equation}
where $\alpha \geq 1.5$ is selected to avoid over-fitting. Note that \eqref{equ:total_order_scaling} corresponds to a \emph{total order} index set, where the sum of the composite univariate polynomial orders in the multivariate expansion is less than or equal to $k$. Such a basis was used in \cite{seshadri2018turbomachinery}, resulting in a total of 351 coefficients (setting $d=25$ and $k=2$ in \eqref{equ:total_order_scaling}) that needed to be estimated. Once obtained, computing gradients of this global quadratic is trivial, resulting in the following estimate of the covariance matrix
\begin{equation}
\mK \approx \frac{4}{3}\mA^{2}+ \vc \vc^{T}
\label{equ:K}
\end{equation}
from which $\mW_{1}$ can be readily obtained following the eigendecomposition of this symmetric matrix. Note that in \eqref{equ:K}, we have assumed a uniform measure
\begin{equation}
\rho \left( \vx \right) = \frac{1}{2^{d}}.
\end{equation}

\subsection{Results on the three blades}
Design of experiments (DoE) were carried out on the three blades, with the total number of CFD evaluations shown in Table \ref{table:3blades}. Given the potential for noise in fitting a global quadratic to the data, some oversampling is necessary. Each design of experiment involved generating $N$ uniformly distributed random samples, i.e., Monte Carlo samples, within the hypercube 
\begin{equation}
\mathcal{X} \subset [-1, 1]^{25} \; \; \; \; \text{where} \; \; \; \; \vx_{i} \in \mathcal{X} \; \; \; \text{for} \; i=1, \ldots, N. 
\end{equation}
Each sample $\vx_{i} \in \mathcal{X}$ was then scaled based on the minimum and maximum values of the design parameters within the CFD workflow. Data-sets for these blades can be accessed at \url{https://github.com/psesh/turbodata} and were used to determine their active subspace via the global quadratic model. 

\begin{figure}
\centering
\begin{subfigmatrix}{2}
\subfigure[]{\includegraphics[]{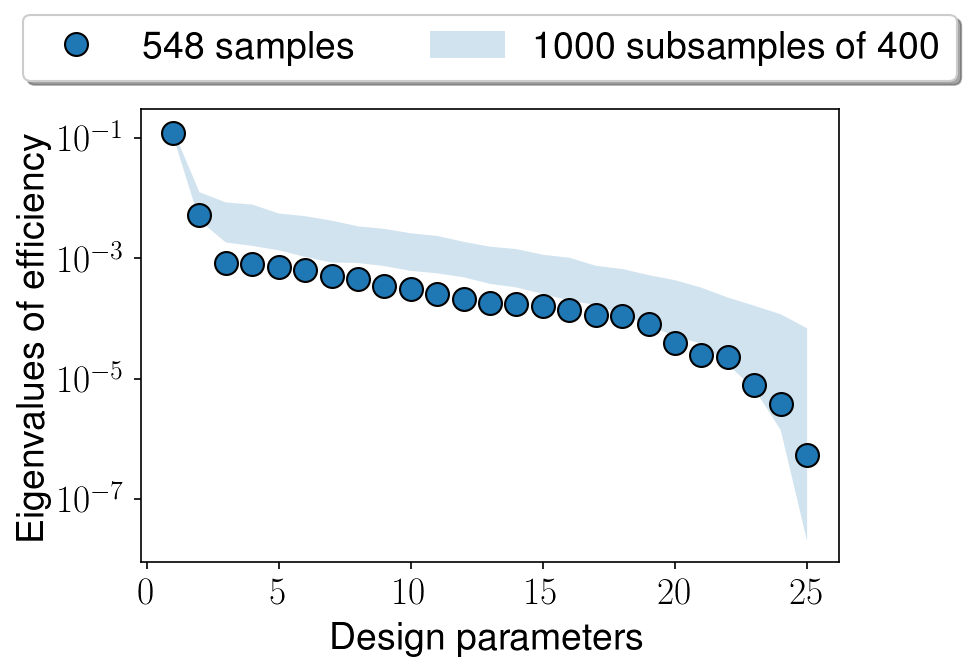}}
\subfigure[]{\includegraphics[]{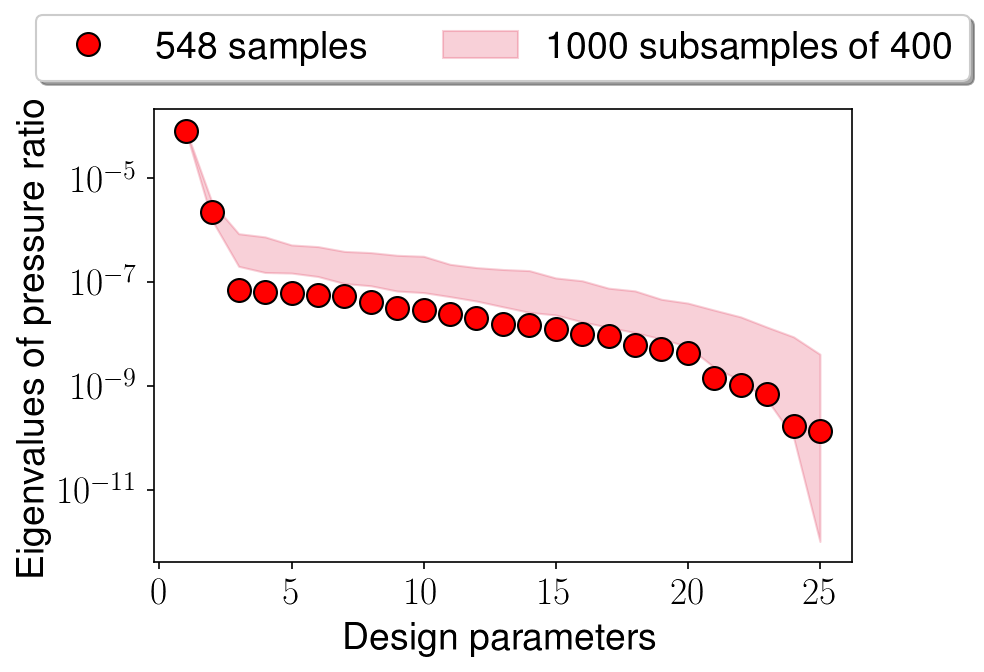}}
\subfigure[]{\includegraphics[]{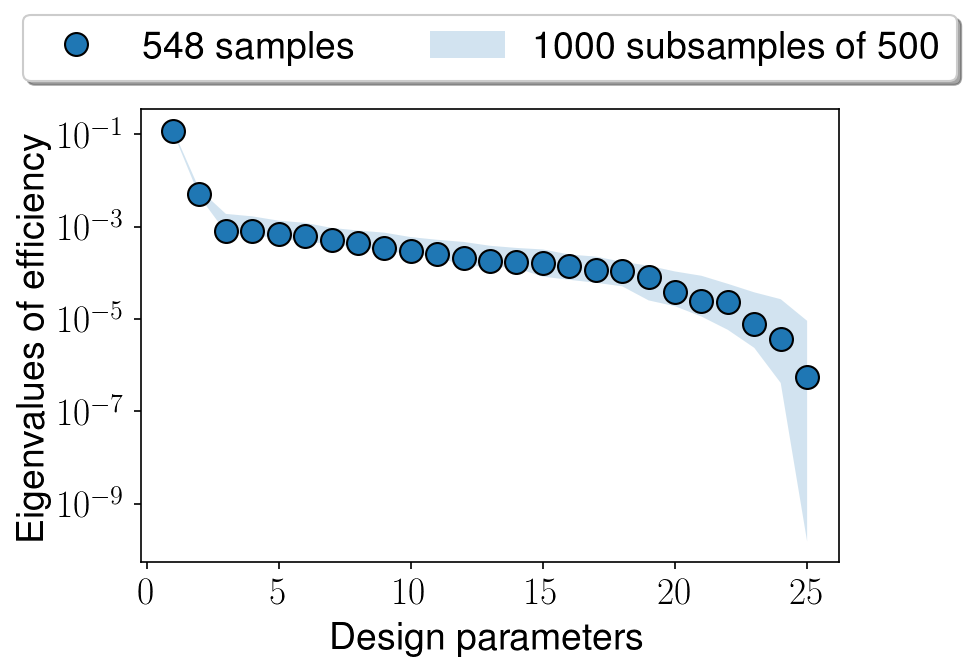}}
\subfigure[]{\includegraphics[]{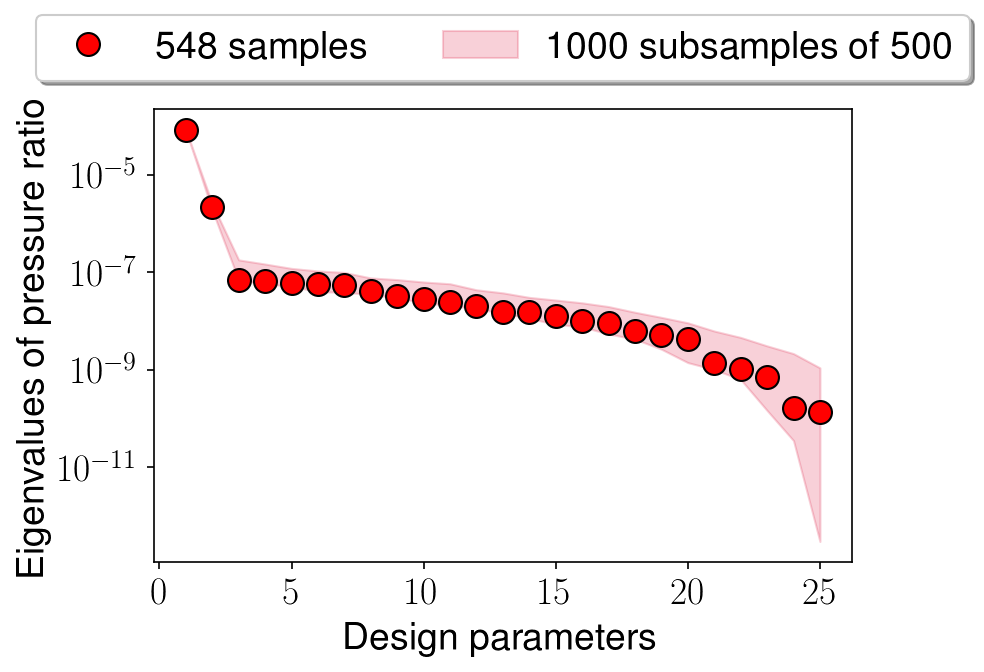}}
\end{subfigmatrix}
\caption{Eigenvalues of the covariance matrix $\mK$ for Blade A for: (a) efficiency with 400 samples; (b) pressure ratio with 400 samples; (c) efficiency with 500 samples; (d) pressure ratio with 500 samples. Shaded regions indicate minimum and maximum eigenvalues from 1000 random repetitions.}
\label{cov_eigen_values_1}
\end{figure}

The eigenvalues for the covariance matrix $\mK$ for these two objectives across for Blade A is shown in Figs.~\ref{cov_eigen_values_1}. These plots are useful to ascertain the dimensionality of the active subspace; Constantine \cite{constantine2015active} suggests that we fix $m$ based on the appearance of large gaps (see page 37 in \cite{constantine2015active}) in the eigenvalues. Typically, this is done in conjunction with a parametric bootstrap with 500-1000 replicates. Now, as the number of CFD evaluations in our three cases is far too low for a parametric bootstrap, we adopt a different approach to probe the accuracy of the active subspace. We compare the eigenvalues obtained using all 548 samples with those obtained by selecting 400 random samples without duplicates. In total a 1000 subsamples are selected; the range of obtained eigenvalues is reported in Figs.~\ref{cov_eigen_values_1}(a) and (b) for the two objectives. In Figs.~\ref{cov_eigen_values_1}(c) and (d) we increase this to 500 subsamples. 

\begin{figure}
\centering
\begin{subfigmatrix}{2}
\subfigure[]{\includegraphics[]{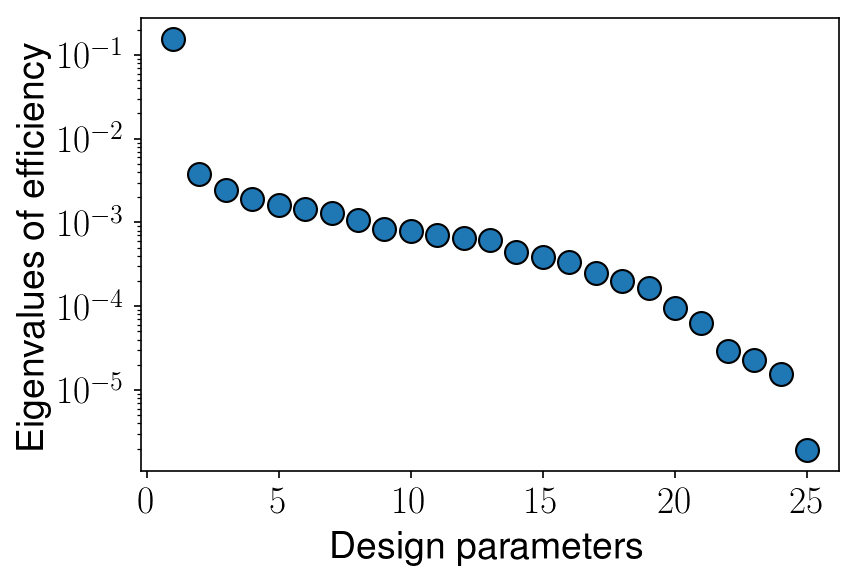}}
\subfigure[]{\includegraphics[]{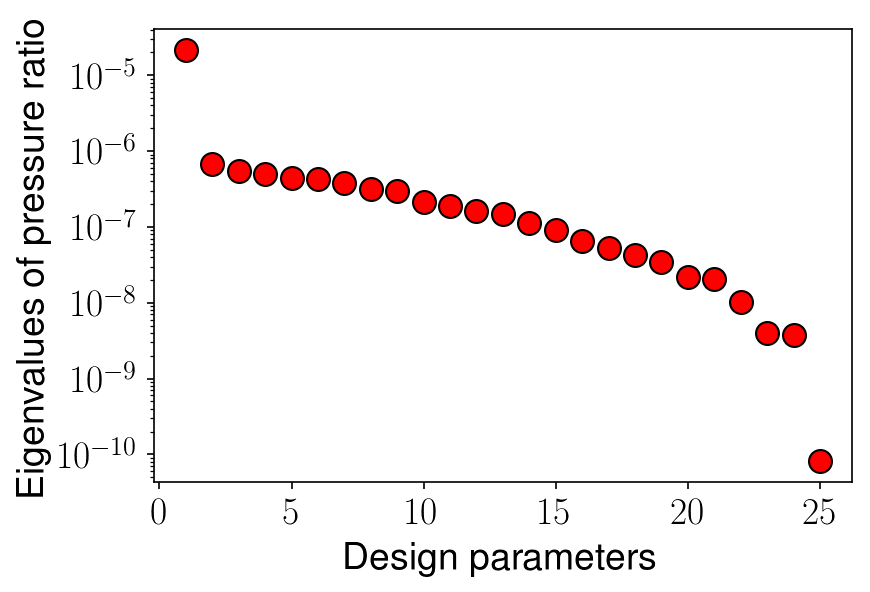}}
\subfigure[]{\includegraphics[]{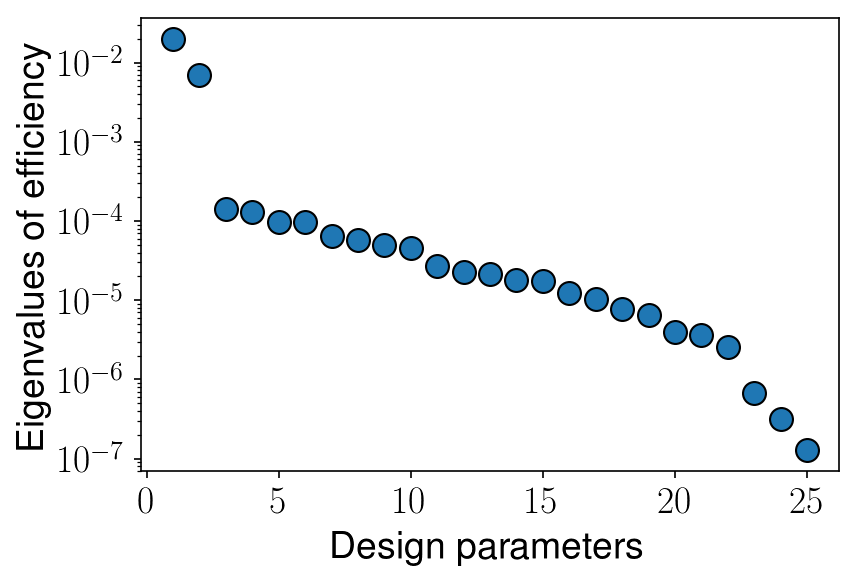}}
\subfigure[]{\includegraphics[]{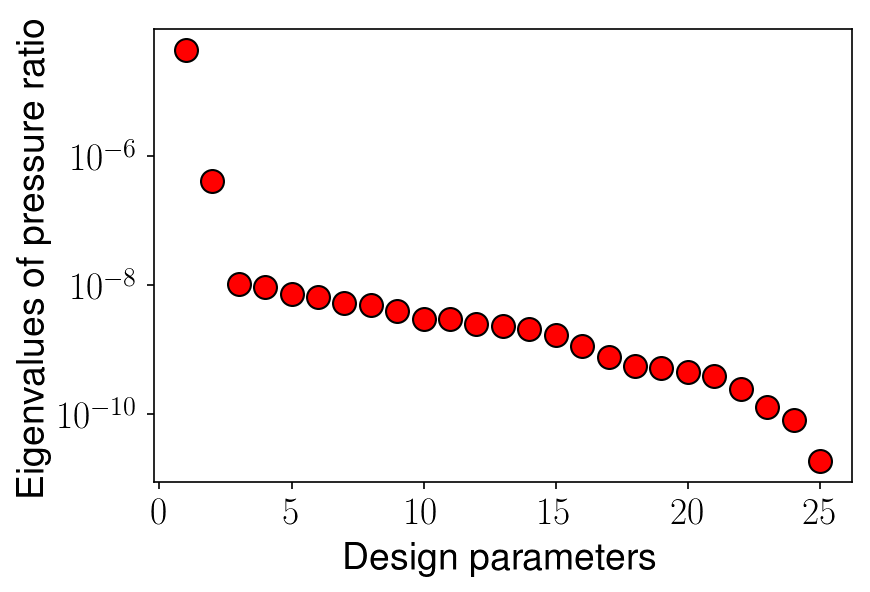}}
\end{subfigmatrix}
\caption{Eigenvalues of the covariance matrix $\mK$ for Blade B for: (a) efficiency (b) pressure ratio, and for Blade C for (c) efficiency (d) pressure ratio.}
\label{cov_eigen_values_2}
\end{figure}

\begin{figure}
\centering
\includegraphics[scale=0.35]{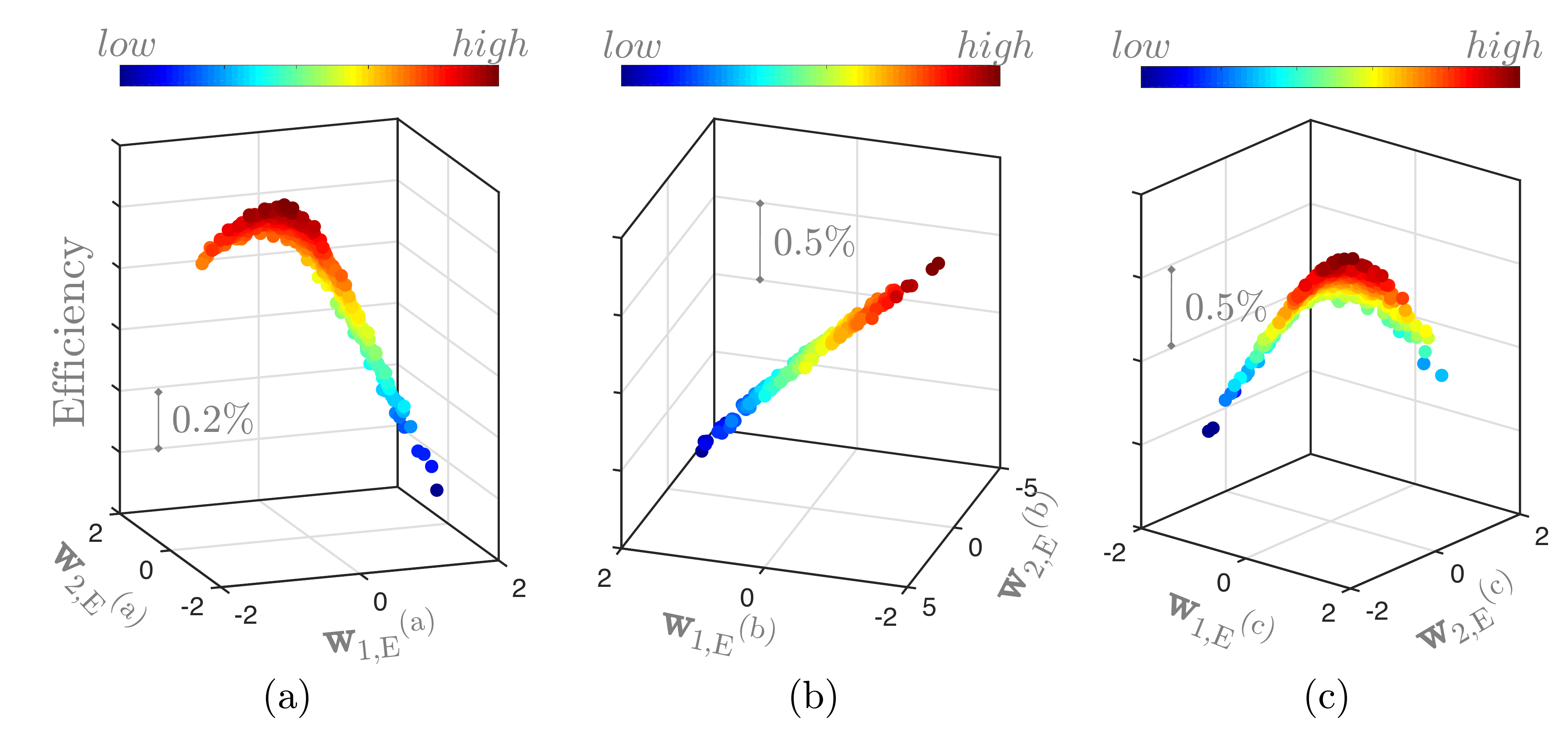}
\caption{Sufficient summary plots for efficiency values for blades (a) A, (b) B and (c) C,  plotted on their active subspaces for efficiency.}
\label{threeblades_eff}
\end{figure}

\begin{figure}
\centering
\includegraphics[scale=0.35]{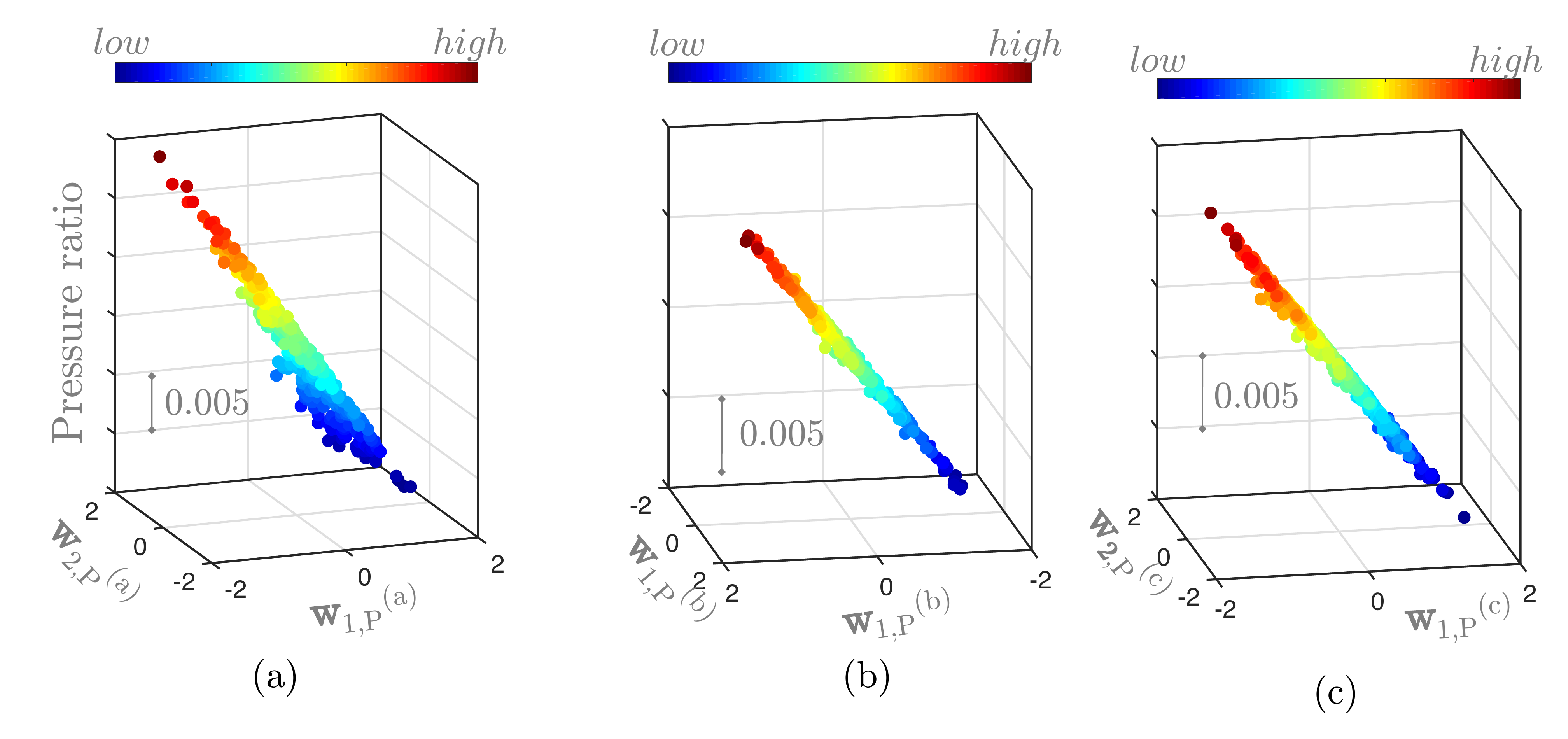}
\caption{Sufficient summary plots for pressure ratio values for blades (a) A, (b) B and (c) C,  plotted on their active subspaces for pressure ratio.}
\label{threeblades_pr}
\end{figure}

A similar study was undertaken for the other two blades; for brevity we only plot the eigenvalues obtained from using all the samples in Fig.~\ref{cov_eigen_values_2}. Based on these results, for both objectives, for Blades A and C, we can set $m=2$ and for Blade B, $m=1$. Sufficient summary plots for the three studies are shown in Figs.~\ref{threeblades_eff} and \ref{threeblades_pr} for the efficiency and pressure ratio objectives respectively. For plotting convenience, we have set $m=2$ for Blade B too.

\begin{table}
\caption{\label{table:3blades}Number of CFD evaluations in DoE processes.}
\begin{center}
  \begin{tabular}{llll}
    \hline
    Blade & No. of CFD evaluations & Description \\  \hline
    A & 548 & High-speed fan\\
    B & 381 & Low-speed fan\\ 
    C & 547 & High-speed fan\\ \hline
  \end{tabular}
\end{center}
\end{table}

In these figures setting $m=2$ yields $\mW_{1} \in \mathbb{R}^{25 \times 2}$. The horizontal axes in each of the six sub-figures here denote the first and second eigenvectors in $\mW_{1}$, i.e.
\begin{equation}
\mW_{1}=\left[\begin{array}{cc}
\vw_{1} & \vw_{2}\end{array}\right]
\end{equation}
The superscripts in the sub-figures denote the blade---i.e.~(a), (b) or (c)---and the subscripts denote the objective---i.e.~$E$ for efficiency and $P$ for pressure ratio. The markers in Figs.~\ref{threeblades_eff} and \ref{threeblades_pr} are then obtained by projecting each $\vx_{i}$ onto these coordinates, with the vertical axis showing the values of the efficiency and pressure ratio.

It is apparent that on their respective subspaces, Blades A and C admit a quadratic variation in efficiency and a largely linear variation in pressure ratio. What is interesting is that for Blade B, which is a low-speed blade, the trend in efficiency is relatively linear. Further examination of these results reveals that the active subspace imparts a change in blade camber to the geometries and, depending on the amount of camber incorporated, the impact of the shock can either be advantageous or debilitating. For Blade B, given its relatively lower operating Mach number, no quadratic trend in efficiency is observed.

\begin{figure}
\centering
\includegraphics[scale=0.35]{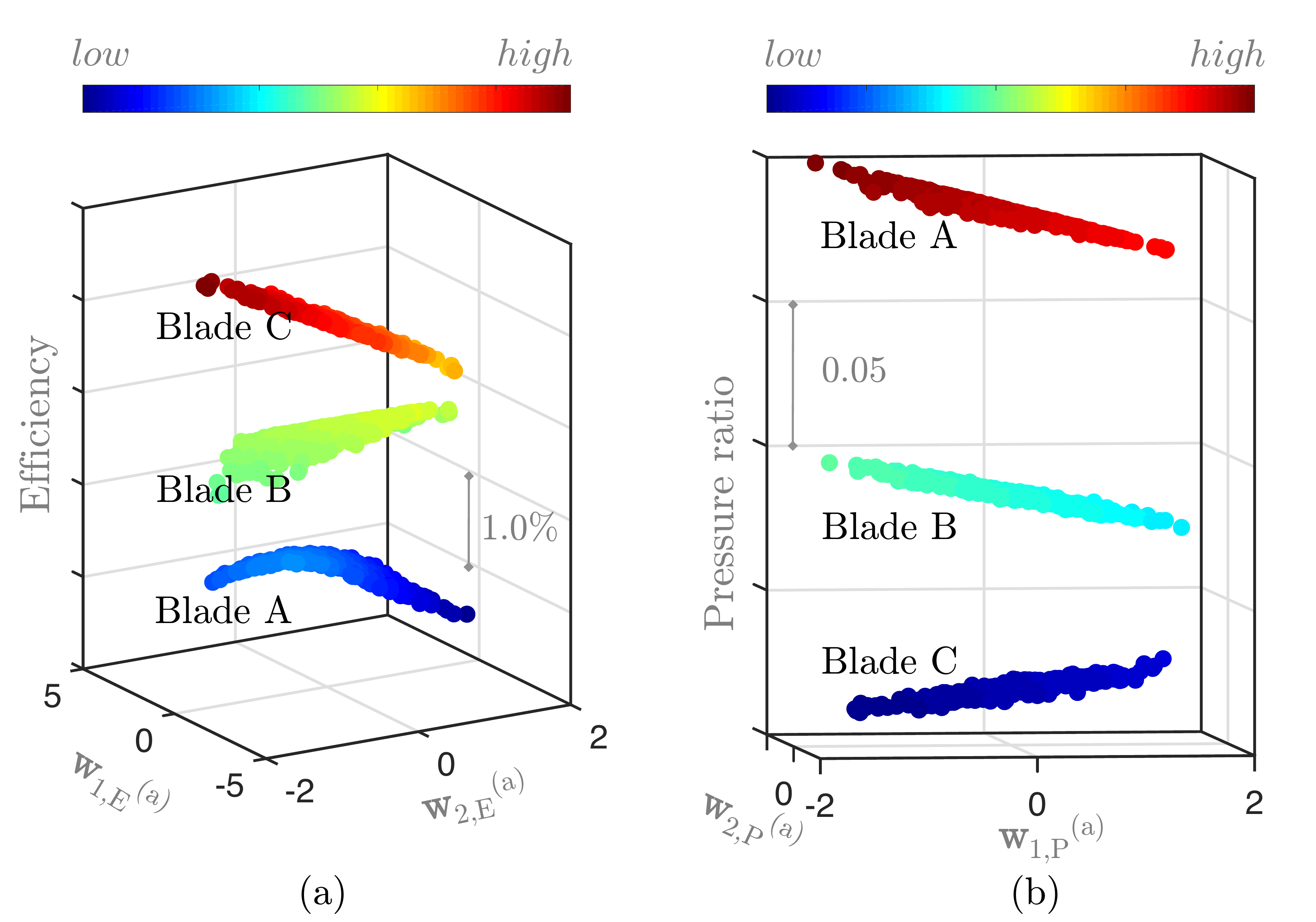}
\caption{Sufficient summary plots for (a) efficiency and (b) pressure ratio values, for blades A, B and C plotted on the active subspace associated with Blade A.}
\label{threeblades_combined}
\end{figure}

\begin{figure}
\centering
\includegraphics[scale=0.35]{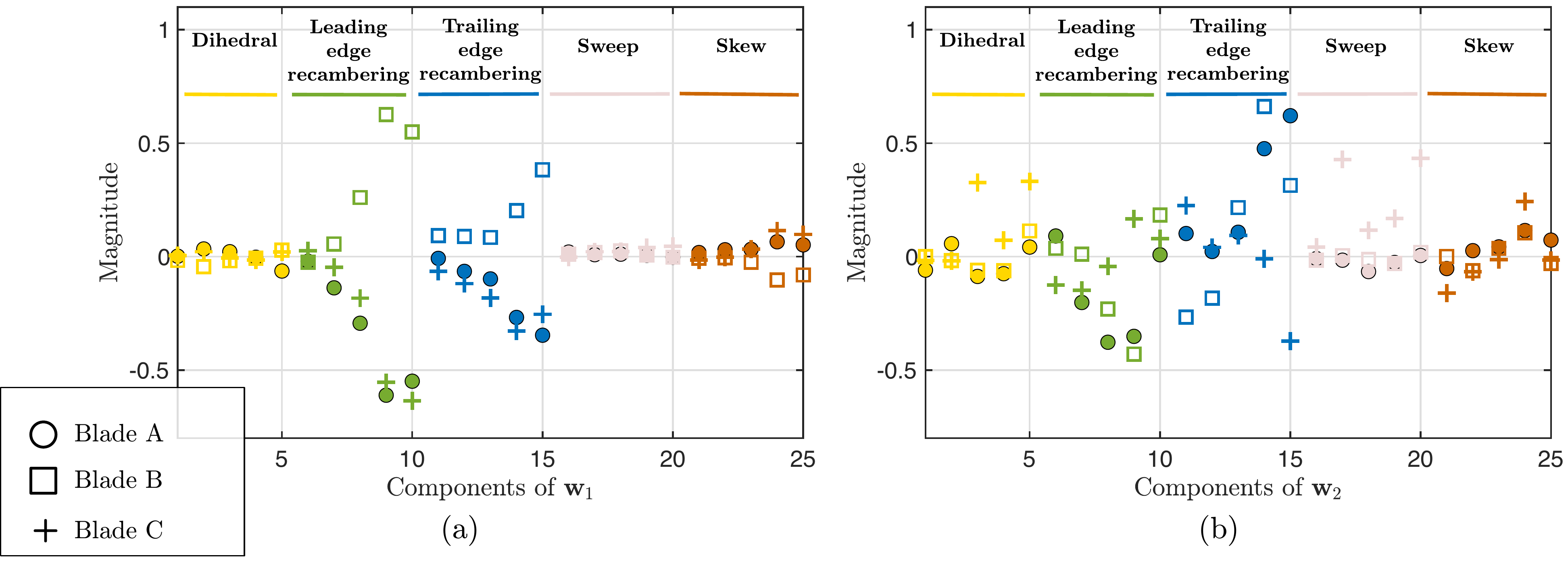}
\caption{Components of the (a) first and (b) second eigenvectors of the covariance matrix for the efficiency active subspace for all three blades.}
\label{threeblades_combined_eff}
\end{figure}

The results above attest to the notion that the active subspaces are driven by underlying physical phenomena and can be found across various blade geometries---note that the nominal geometries are different---at different operating conditions. But can we \emph{re-use} the active subspaces obtained for one blade on another? In Fig.~\ref{threeblades_combined}, we plot the efficiency and pressure ratio values of Blades B and C on the active subspace associated with Blade A. This sufficient summary plots clearly captures the trends we observed above and demonstrates that a new blade geometry with the same parameterization can leverage the subspace discovered from another blade, even though each blade's active subspace is different (see Fig.~\ref{threeblades_combined_eff}). This implies that for a new blade, we may not require as many CFD evaluations as for the three blades above as we can utilize the subspace previously computed.

We end this section with a remark on the computational cost of approximating $\mW$ and, by association, its active subspace. The cost of a global quadratic model scales poorly with dimension, motivating a different strategy to effectively compute the active subspace: for a 50-dimensional problem the number of coefficients that need to be estimated rises to 1326; for a 100-dimensional problem to 5151, etc. 

%% file: sec-3.tex
\section{Statistical Sufficient Dimension Reduction}
\label{sec:two}
In this section, we revisit well-known statistical strategies for sufficient dimension reduction (SDR), in the hope of addressing the computational cost of the quadratic approach detailed previously. It should be noted that these statistical techniques are rooted in classical regression, while ideas within the remit of active subspaces are rooted in the field of simulation-driven modeling. The first difference between these two schools of thought is that statistical methods often assume a zero-mean random error in the model. However, for computer experiments, which are usually deterministic, there is no such error. The second difference is that the joint probability density function of the input samples in computer experiments is known. We point the interested reader to Sect.~2.2 of \cite{seshadri2018dimension} for further details on these differences and to the work of Glaws et al. \cite{glaws2018inverse, glaws2017inverse}, who provide a theoretical and computational framework for utilizing two well-known SDR techniques with computer models. 

With this preface, we discuss four classical dimension reduction approaches: sliced inverse regression (SIR) \cite{li1991sliced}, sliced average variance estimation (SAVE) \cite{cook1991comment}, principal Hessian directions (pHd) \cite{li1992principal} and contour regression (CR) \cite{li2005contour}. Our exposition of these approaches is paired with their application to the DoE results on Blade A, and closely follows the notation adopted in Glaws et al.~\cite{glaws2018inverse}.

\subsection{Theory and computational approaches}
As a necessary pre-processing step, the inputs to all the aforementioned methods need to be standardized, aimed at the reduction of multicollinearity (see Sect.~5.3 in \cite{faraway2016linear}). Analogous to ideas for computing the \emph{active subspace}, SDR methods require the construction of a covariance matrix, the dominant eigenvectors of which define directions within the input parameter space that are important. While for active subspaces our metric of \emph{importance} was related to $\nabla_{\vx} f$, SDR techniques adopt different metrics. In SIR, for instance, the covariance matrix is given by
\begin{equation}
\tilde{\mK}=\frac{1}{N}\sum_{s=1}^{S}N_{s} \boldsymbol{\mu}_{s}  \boldsymbol{\mu}_{s}^{T}
\label{equ_SIR}
\end{equation}
where
\begin{equation}
\boldsymbol{\mu}_{s}=\frac{1}{N_{s}}\sum_{j\in\mathcal{J}_{s}} \vx_{j}
\end{equation}
Here $s=1, \ldots, S$ represents the slices into which the output values $f \left(  \vx_{i} \right)$ for $i=1, \ldots, N$ must be uniformly\footnote{Li \cite{li1991sliced} relaxes this requirement in Sect.~6 of his paper, i.e.~each slice need not have the same number of samples.} placed, such that each slice contains $N_s$ samples, i.e.~$N_{s}=\left|\mathcal{J}_{s}\right|$, where 
$\mathcal{J}_{s} \in [ 1, \ldots, N ]$. The term $\boldsymbol{\mu}_{s}$ is the sample mean of the input parameters in the $s$-th slice, making SIR a method based on the first moment only. Additionally, due to its dependence on the first moment, SIR fails to handle symmetric functions. 

Once the covariance matrix in Eq.~\eqref{equ_SIR} has been constructed, its first few eigenvectors can inform important directions. One apparent advantage of SIR is that it does not require gradient information. Further, while the number of slices $S$ may alter the asymptotic variance of the output, the effect may not be significant (see Remark 4.3 in \cite{li1991sliced}). 

A similar slicing strategy is adopted in SAVE, a technique based on both the mean and the variance. Here the covariance matrix has the form
\begin{equation}
\tilde{\mK}=\frac{1}{N}\sum_{s=1}^{S}N_{s}\left( \boldsymbol{I}- \boldsymbol{\Omega}_{s} \right)^{2}
\label{equ_SAVE}
\end{equation}
where sample covariances are given by
\begin{equation}
\boldsymbol{\Omega}_{s}=\frac{1}{N_{s}-1}\sum_{j\in\mathcal{J}_{s}}\left( \vx_{i}- \boldsymbol{\mu}_{s}\right)\left( \vx_{i}- \boldsymbol{\mu}_{s}\right)^{T}
\end{equation}
and where $\boldsymbol{I}$ is the identity matrix. 

The next SDR approach we introduce is pHd, a technique that does not require slices, differentiating it from SIR and SAVE.  As Li \cite{li1992principal} states, \emph{``[pHd] begins with the observation that the eigenvectors for the Hessian matrices of the regression function are helpful in the study of the shape of the regression surface.''} The method tries to locate the principal directions along which the regression surface, on average, exhibits the greatest curvature variation. Thus, its covariance matrix seeks to approximate the average Hessian of the function, i.e. $\mathbf{K} = \mathbb{E}[\nabla^2 f(\vx)]$. To achieve this, Li \cite{li1992principal} proposes the following approximation
\begin{equation}
\tilde{\mK}=\frac{1}{N}\sum_{i=1}^{N}\left(f\left(\vx_{i}\right)-\bar{f}\left( \vx\right) \right)\left(\vx_{i}-\bar{\vx}\right)\left(\vx_{i}-\bar{\vx}\right)^{T},
\end{equation}
where 
\begin{equation}
\bar\vx=\frac{1}{N}\sum_{i=1}^{N}\vx_i
\end{equation}
is the sample mean for the inputs and
\begin{equation}
\bar{f} \left( \vx\right) =\frac{1}{N}\sum_{i=1}^{N}  f \left( \vx_{i}\right)
\end{equation}
is the sample mean for the outputs. It should be noted that, in general, pHd is less stable than SIR, due to its dependence on second-order moments. 

The last SDR method we shall review here is CR \cite{li2005contour}, where empirical directions are defined as the vectors
\begin{equation}
\vx_i-\vx_j \; \; \forall \; \; 1 \leqslant i\neq j \leqslant N
\end{equation}
The method extracts a subset of empirical directions with small variations in $f \left(\vx \right)$. Then the directions with largest variations in $f \left(\vx \right)$ become estimates for the dimension reducing subspace. The covariance matrix for CR is given by
\begin{equation}
\hat{ \mK}(c)=\frac{2}{N(N-1)}\sum_{1\leqslant i\neq j \leqslant N}(\vx_j-\vx_i)(\vx_j-\vx_i)^{\rm T}  \mathcal{I} \left( \lvert f \left(\vx_j \right)  -  f \left(\vx_i \right)  \rvert \leqslant c  \right)
\end{equation}
where
\begin{equation}
\mathcal{I} \left(   \lvert f \left(\vx_j \right)  -  f \left(\vx_i \right)     \rvert \leqslant c  \right)=
\begin{cases}
1 \; \; \;\text{for } \lvert  f \left(\vx_j \right)  -  f \left(\vx_i \right)     \rvert \leqslant c\\
0 \; \; \;\text{for } \lvert   f \left(\vx_j \right)  -  f \left(\vx_i \right)   \rvert > c\\
\end{cases}
\end{equation}
and $c$ is a user-defined tolerance. The dimension reducing subspaces are then found by selecting eigenvectors corresponding to the smallest eigenvalues. One drawback of CR is that the tolerance $c$ needs to be chosen. 

\subsection{Results on Blade A}
\begin{figure}
\centering
\includegraphics[scale=0.5]{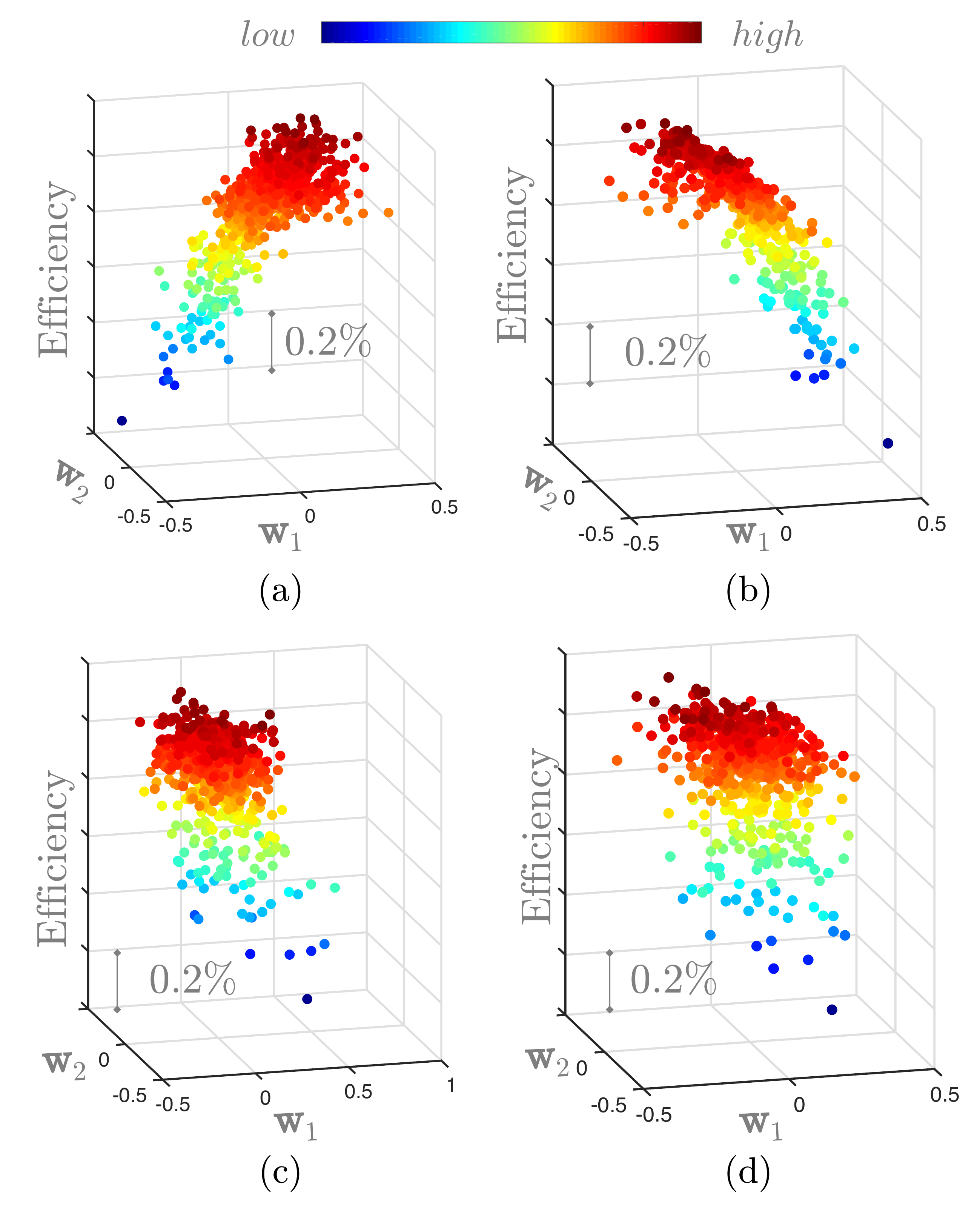}
\caption{Sufficient summary plots for the Blade A data-set for: (a) SIR; (b) SAVE; (c) pHd; (d) CR.}
\label{sdr_a}
\end{figure}

In what follows we apply the four SDR methods discussed above on the data-set associated with Blade A, focusing only on the efficiency objective. Recall that our overarching aim in this section is to ascertain whether SDR rooted methods can identify the dimension reducing subspaces---similar to those shown in Sect.~\ref{sec:one}---on a more parsimonious computational budget. To this end, we randomly select 200 input-output ($\vx_{i}, f \left( \vx_{i} \right) $) pairs from the computed 548 pairs. These 200 pairs are then given as inputs to SIR, SAVE, pHd and CR. The computed subspaces obtained using these techniques are shown in Fig.~\ref{sdr_a} in the form of sufficient summary plots. Once the subspaces have been computed, we project the remaining 348 samples as well. 

\begin{figure}
\centering
\includegraphics[scale=0.5]{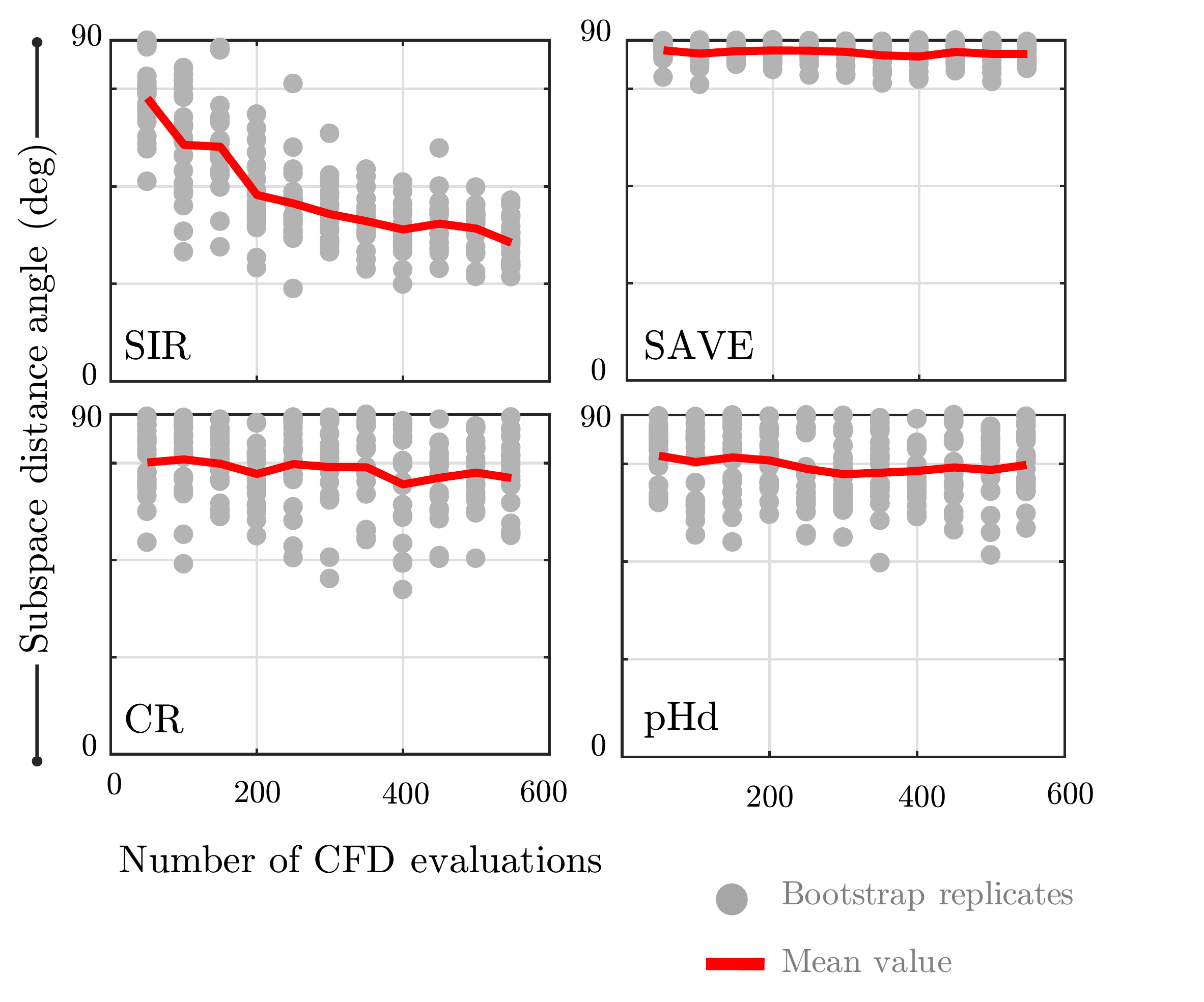}
\caption{Subspace distance angles between the SDR and active subspace. The gray markers show the outcomes of 20 bootstrap replicates, while the red line shows their mean values.}
\label{sdr_b}
\end{figure}

To understand the effect of the number of samples (and our random subsampling), we introduce a convergence metric
\begin{equation}
\phi=\arcsin \left(   \left\Vert   \tilde{\mW}_{1}\tilde{\mW}_{1}^{T}-\mW_{1}  \mW_{1}^{T} \right\Vert _{2} \right)
\end{equation}
defined as the angle between the subspace approximated by the various SDR methods, $\tilde{\mW}_{1}$, and the subspace obtained through the quadratic model via the eigendecomposition of Eq.~\eqref{equ:K} yielding $\mW_{1}$. We report the change in $\phi$ as a function of the number of subsamples for the four different SDR strategies in Fig.~\ref{sdr_b} with 20 bootstrap replicates for a given number of samples. Our study shows that SAVE, CR and pHd fail to identify the underlying structure in the data, while SIR obtains an approximation that is relatively closer to that of the quadratic model. 

We remark here that although we have used active subspaces as a basis\footnote{Literally.} for comparison, we cannot state that the active subspace is \emph{optimal}. What we can state is that the active subspaces offers a subspace where statements on the gradient of the residual $\left\Vert f\left( \vx\right)-g\left(\mW_{1}^{T} \vx\right)\right\Vert _{2}^{2}$ can be made (see Definition 1 of \cite{constantine2017near}). It can serve as an initial solution, if not local minima, to the problem of approximating a high dimensional function over an $m-$dimensional subspace, but multiple other subspaces can exist. Thus, care should be taken when interpreting Fig~\ref{sdr_b} in the absence of Fig.~\ref{sdr_a} as the latter is just as important.

%% file: sec-5.tex
\section{Polynomial Variable Projection for Dimension Reduction}
\label{sec:three}
The SDR methods applied to our problem have clearly been unable to identify similar structures to those we observed when using the active subspaces global quadratic model. Once again, motivated by the need to parsimoniously obtain dimension reducing subspaces, we study a new idea proposed by Hokanson and Constantine \cite{hokanson2017data}, which frames the subspace discovery problem as a non-linear least squares problem; our notation closely follows theirs.

\subsection{Theory and computational approaches}
Let us start by redefining Eq.~\eqref{equ_active} as
\begin{align}
\begin{split}
f \left( \vx \right) & \approx g \left( \mU^{T} \vx \right) \\
& = \sum_{i=1}^{m}\alpha_{i} \boldsymbol{\psi}_{i} \left( \mU^{T} \vx\right)
\end{split}
\end{align}
where $g$ is a polynomial in $n<<m$ dimensions with a maximum order of $p$ with cardinality $\rho$, and where $\mU \in \mathbb{R}^{m \times n}$ is an orthogonal matrix with $m=25$ for our studies. The matrix $\mU$ belongs to the Grassman manifold of $n$-dimensional subspaces in $\mathbb{R}^{m}$. We express $g$ as a linear combination of coefficients $\alpha_{i}$ for $i=1, \ldots, n$  and multivariate $l_2$ orthogonal polynomial basis terms $\boldsymbol{\psi}_{i}$. These multivariate polynomials are products of composite univariate polynomials $\psi$
\begin{align}
\begin{split}
 \boldsymbol{\psi}_{\boldsymbol{q}}\left( \mU^{T} \vx\right) & =\prod_{j=1}^{n}\psi_{q_{j}}   \left(\vu_{j}^{T} \vx \right) \\
\Rightarrow \boldsymbol{\psi}_{\boldsymbol{q}}\left(  \vy \right) & =\prod_{j=1}^{n}\psi_{q_{j}}  \left(  y^{\left( j \right) }    \right) 
\label{equ_poly}
\end{split}
\end{align}
where the superscripts are used to designate the relevant dimension. Here the symbol $\boldsymbol{q} = (q_1, \ldots, q_n) \in \mathbb{N}^{n}$ is a finite set of indices that belongs to a multi-index set $\mathcal{I}$. In this paper, we use both isotropic tensor product and isotropic total order bases, the difference between them being the number of basis terms, i.e.~the \emph{cardinality} of the index set. For a maximum degree of $p$, the former is given by the rule $\text{max} \left( q_j \right) \leq p$, while for the total order basis $\sum_{i=1}^{n}q_{i}\leq p$ (see Sect.~1.2 in \cite{seshadri2017effectively} for further details). The notation $\vu_{j}$ denotes the $j$-th column of $\mU$ and $y^{ \left( j\right) } = \vu^{T}_{j} \vx$ for $j=1, \ldots, n$, represents the projected coordinates on the space spanned by the columns of $\mU$. Gradients of Eq.~\eqref{equ_poly} with respect to $\vy = \mU^{T} \vx$ can be easily computed via
\begin{equation}
\frac{\partial \boldsymbol{\psi}_{q}}{\partial y^{(l)}}=\frac{d\psi_{ql}}{dy^{(l)}}\prod_{j=1,j\neq l}\psi_{qj}\left(y^{\left(j\right)}\right).
\label{equ:grads}
\end{equation}
These gradients will be required in subsequent computations.

Our main objective in this section is to solve the optimization problem
\begin{equation}
\underset{ \stackrel{\alpha_{i}, \; i=1,\ldots,n}{\text{range}\; \mU \in \mathbb{G}(n, \mathbb{R}^m) } }{ \text{minimize} }
 \; \; \left\Vert f\left(\vx\right)-   \sum_{i=1}^{n}\alpha_{i} \boldsymbol{\psi}_{i} \left( \mU^{T}x\right)       \right\Vert^{2}_{2}
\label{equ:vapro}
\end{equation}
which can be expressed as
\begin{equation}
\underset{\boldsymbol{ \alpha} ,\mU }{\text{ minimize} }\left\Vert \vf- \mP\left(\mU^{T} \vx\right) \boldsymbol{\alpha} \right\Vert _{2}^{2}
\label{equ:vapro2}
\end{equation}
where $\vf=\left[f\left(\vx_{1}\right) \dots f\left(\vx_{K}\right)\right]^{T}$, $\boldsymbol{\alpha} =\left[\alpha_1 \dots \alpha_n \right]^{T}$ and
\begin{equation}
\mP(i, j) = \boldsymbol{\psi}_{j} \left( \mU^{T} \vx_{i} \right)
\label{equ:vapro3}
\end{equation}
Using the notion of variable projection \cite{golub1973differentiation} (also see \cite{golub2003separable} and Sect.~8.4 in \cite{hansen2013least}), one can rewrite Eq.~\eqref{equ:vapro2} as an optimization problem over $\mU$ only
\begin{align}
\begin{split}
\vh  &= \underset{ \mU }{\text{ minimize} }\left\Vert \vf- \mP\left(\mU^{T} \vx\right) \mP^{\dagger}\left(\mU^{T} \vx\right) \vf   \right\Vert _{2}^{2} \\
&= \underset{ \mU }{\text{ minimize} }\left\Vert \left( \mI - \mP\left(\mU^{T} \right) \mP^{\dagger}\left(\mU^{T} \right) \right)  \vf   \right\Vert _{2}^{2} \\
&= \underset{ \mU }{\text{ minimize} }\left\Vert  \vr \left( \mU \right)   \right\Vert _{2}^{2}
\end{split}
\label{final}
\end{align}
where $\vr$ is known as the \emph{projector onto the orthogonal complement of the column space of the matrix} $\mP \left(\mU^{T} \right)$ and the matrix $ \mP^{\dagger}$ is the generalized Moore-Penrose inverse of $\mP$. Now $\vr \in \mathbb{R}^{M}$ where $M$ represents the total number of samples. The first step in the variable projection approach is to formulate both the gradient and the Hessian (see Sect.~3 of \cite{golub2003separable})
\begin{equation}
\frac{1}{2}\nabla\left\Vert \vr\left(\mU\right)\right\Vert _{2}^{2}= \mJ^{T}\left(\mU\right)r\left(\mU\right)
\end{equation}
\begin{equation}
\frac{1}{2}\nabla^{2}\left\Vert \vr \left( \mU\right)\right\Vert _{2}^{2}= \mJ^{T}\left( \mU\right)\mJ( \mU)+\sum_{i=1}^{M}\vr_{i}\left(\mU\right)\nabla^{2} \vr_{i}\left(\mU\right)
\label{equ_hess}
\end{equation}
The Jacobian $\mJ \in \mathbb{R}^{M \times m \times n}$, a tensor, can be written as
\begin{equation}
\mJ \left(:, j, k \right) =-\left(\left(\mI- \mP \mP^{\dagger}\right)\frac{\partial \mP}{\partial \mU_{\left(j,k\right)}} \mP^{-}\right)\vf-\left(\left(\mI- \mP \mP^{\dagger}\right)\frac{\partial \mP}{\partial \mU_{\left(j,k\right)}} \mP^{-}\right)^{T} \vf
\label{equ_jacobian}
\end{equation}
The symbol $\mP^{-}$ denotes the generalized inverse; as stated in \cite{golub1973differentiation}, the full Moore-Penrose pseudoinverse is not required, and thus a symmetric generalized inverse, for which the two conditions
\begin{equation}
\left(\mP \mP^{-} \mP\right)= \mP \; \; \; \;  \text{and} \; \; \; \; \left(\mP \mP^{-}\right)^{T}=\mP \mP^{-}
\end{equation}
hold, will suffice. We also remark here that the power of variable projection stems from the fact that the expression in Eq.~\eqref{equ_jacobian} does not involve gradients of the Moore-Penrose pseudoinverse. To compute the terms in Eq.~\eqref{equ_jacobian}, one simply uses Eq.~\eqref{equ:grads} with Eqs.~\eqref{equ:vapro2} and \eqref{equ:vapro3}. Details on how these terms are computed, along with properties of the Jacobian are provided in \cite{hokanson2017data}. Further, in \cite{hokanson2017data} the authors provide a step-by-step Gauss-Newton algorithm---where the Hessian $\mJ^{T}\left( \mU\right)\mJ( \mU)$ in Eq.~\eqref{equ_hess} is approximated---for solving the optimization problem in Eq.~\eqref{final}. Our implementation of their strategy is available in the Effective Quadratures open-source package \cite{seshadri2017effective}.

\subsection{Results on Blade A}
\begin{figure}
\centering
\includegraphics[scale=0.5]{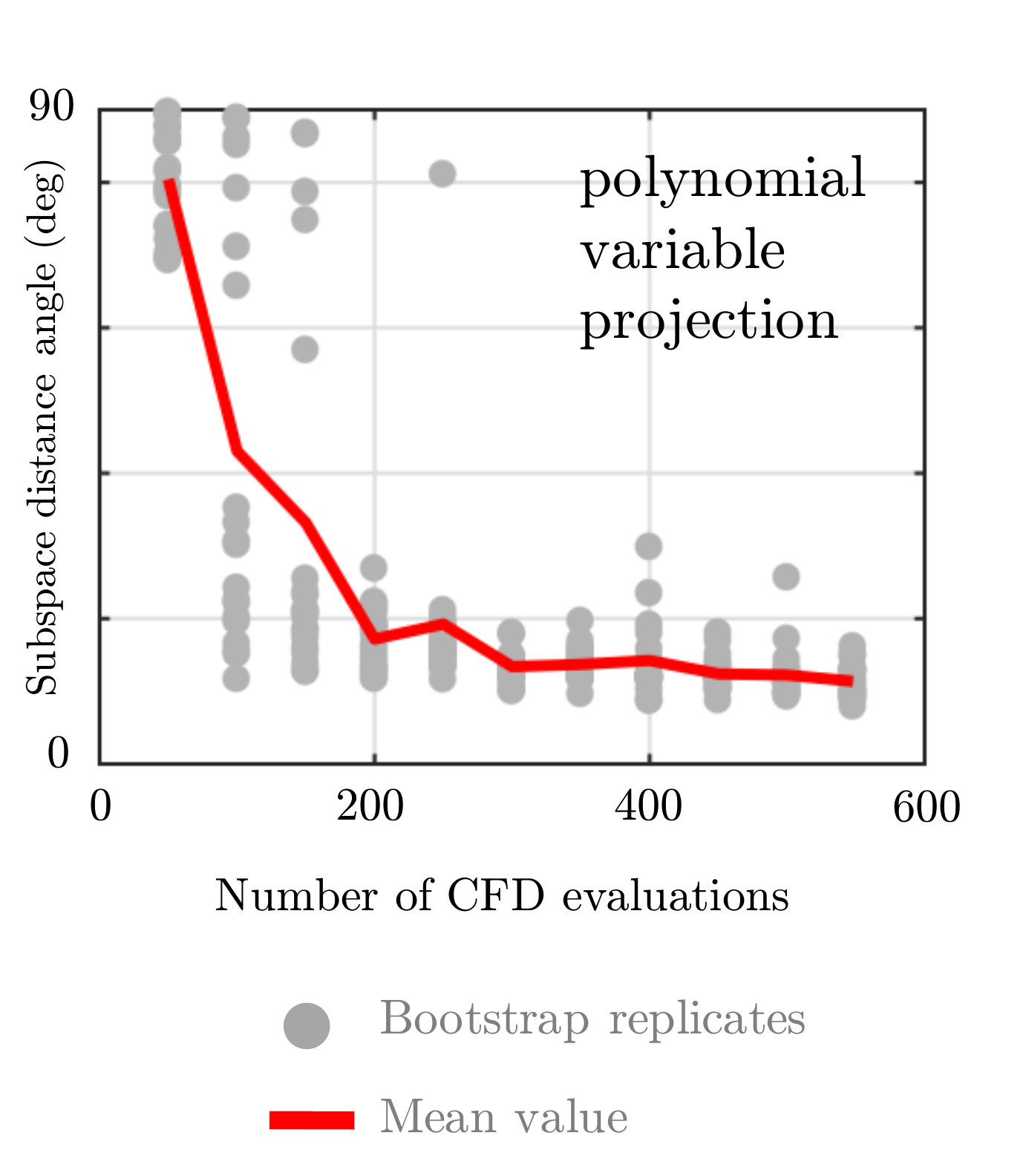}
\caption{Subspace distance angles between the variable projection and active subspace. The gray markers show the outcome of 20 bootstrap replicates, while the red line shows their mean value.}
\label{vp_a}
\end{figure}

We apply variable projection on the data-set associated with Blade A in the same manner as we did with the SDR methods described in Sect.~\ref{sec:two}. Figure \ref{vp_a} shows the results for variable projection with different numbers of samples and their bootstrap replicates. For each experiment, the subspace angle obtained from variable projection is compared with the global quadratic active subspaces approach detailed before. Here we set $p=2$ with a total order basis. Compared to all of the aforementioned techniques, variable projection does yield the lowest angles, indicating that the approximated subspaces are close to the active subspace.

\begin{figure}
\centering
\includegraphics[scale=0.5]{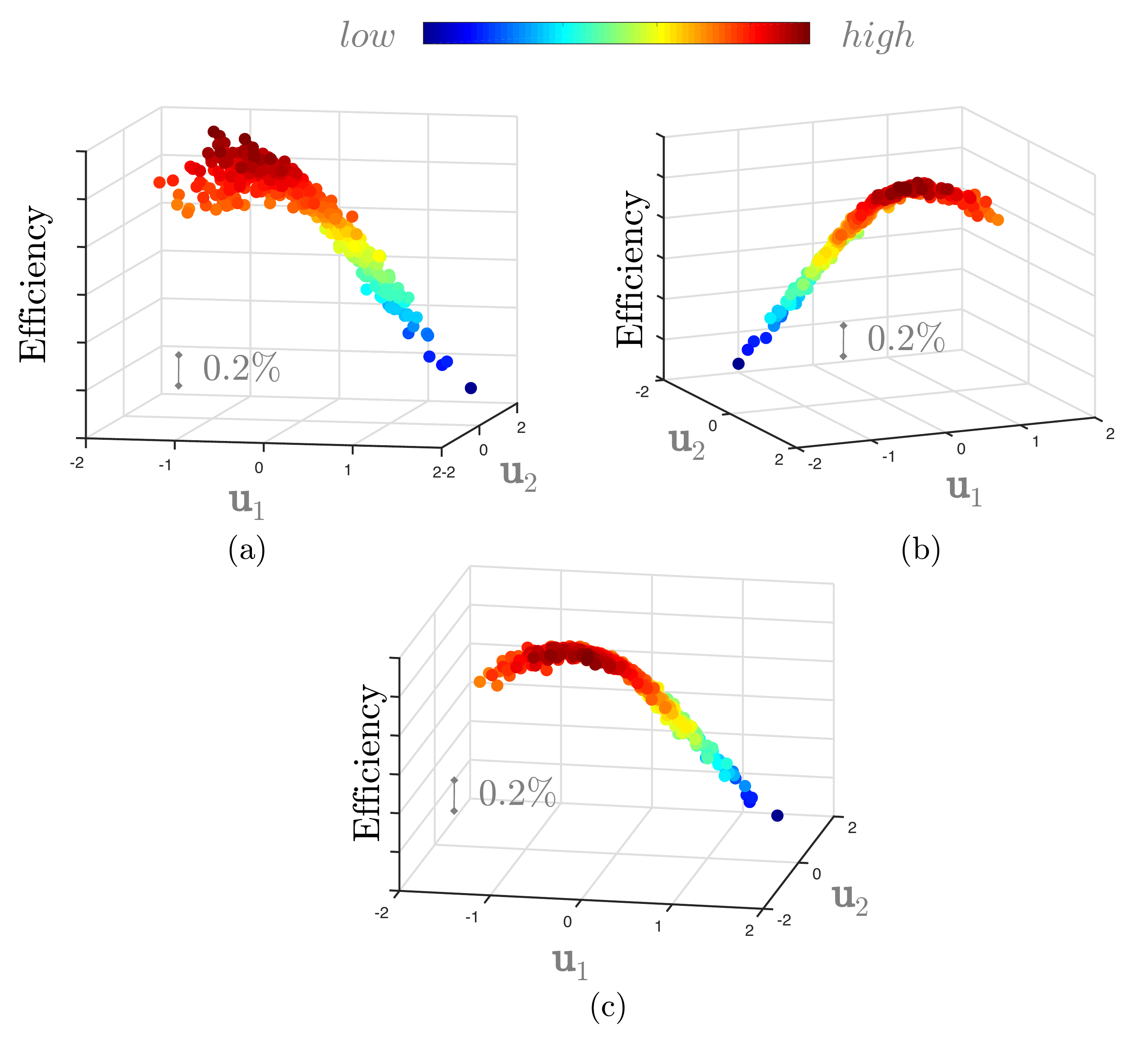}
\caption{Sufficient summary plots of the variable projection (total order degree 2) approach with: (a) 150, (b) 200 and (c) 250 randomly subsampled input-output pairs for Blade A.}
\label{vp_b}
\end{figure}

The results for a few of these experiments are shown in Fig.~\ref{vp_b}. Here sub-figures (a), (b) and (c) show the sufficient summary plots, where the 2D coordinates in the figure correspond to the columns of $\mU = [\vu_{1}, \vu_{2}]$ when evaluating the algorithm with $M=150, 200$ and $250$ samples respectively. These results give us confidence in the utility of variable projection for finding dimension reducing subspaces for our turbomachinery problem. More importantly, the reduced sample count---150 to 200 samples---compared to the cost of a global quadratic model---above 351 samples---represents a significant computational saving. 

\begin{figure}
\centering
\includegraphics[scale=0.5]{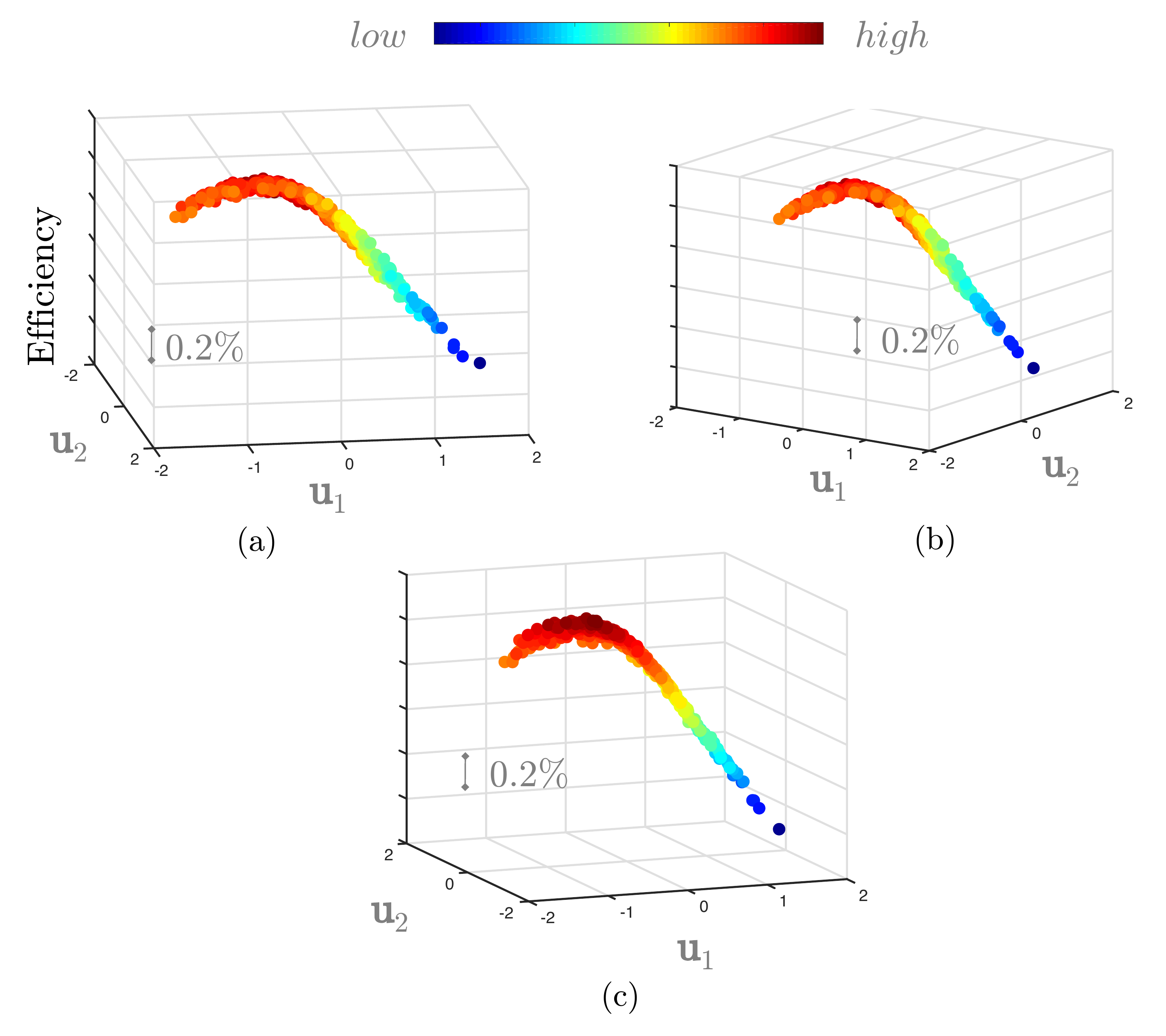}
\caption{Sufficient summary plots of the variable projection (tensor order degree 3) approach with: (a) 150, (b) 200 and (c) 250 randomly subsampled input-output pairs for Blade A.}
\label{vp_b2}
\end{figure}

\begin{figure}
\begin{subfigmatrix}{2}
\subfigure[]{\includegraphics{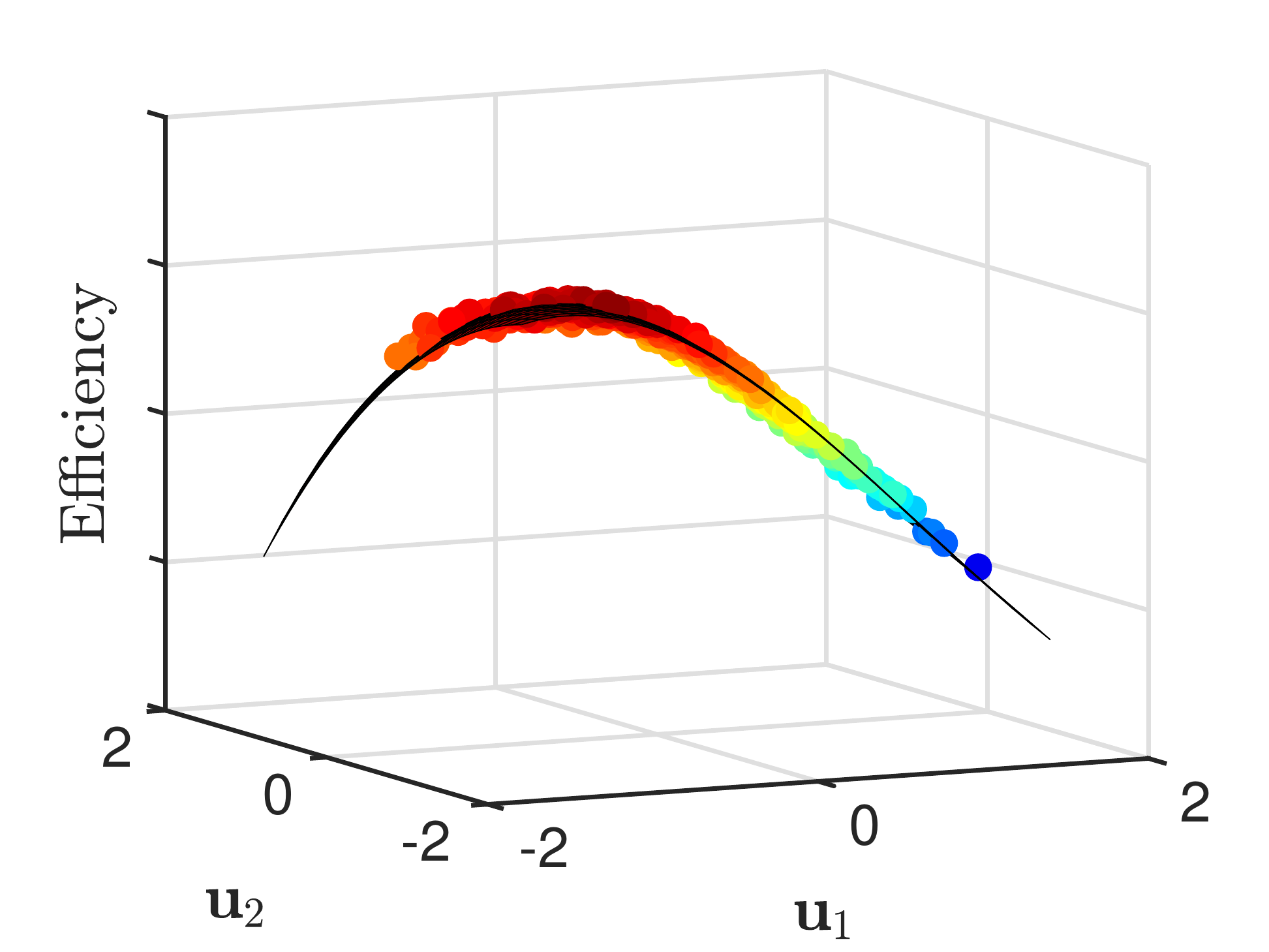}}
\subfigure[]{\includegraphics{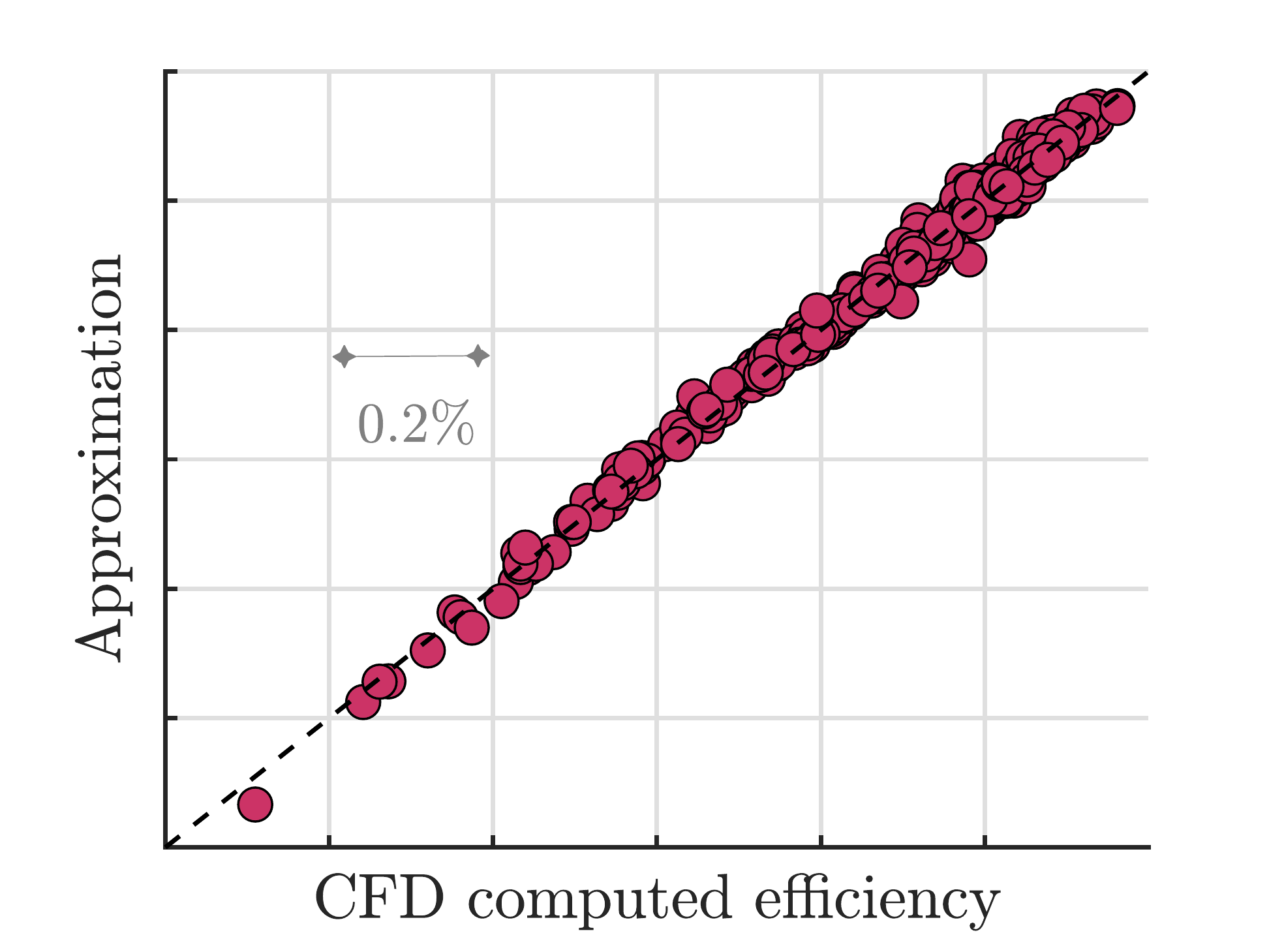}}
\end{subfigmatrix}
\caption{Polynomial (cubic) response surface on the subspace in Fig.~\ref{vp_b2}(c) and a comparison between this polynomial and the CFD results in (b).}
\label{figure_new3}
\end{figure}

In the same vein as before, Fig.~\ref{vp_b2} shows sufficient summary plots for $p=3$ using a tensor order basis; the results are similar, with reduced scatter even when the number of samples used is 150. In Fig.~\ref{figure_new3}(a) we show the polynomial fit for the result in Fig.~\ref{vp_b2}(c) and compare the values of this approximation and the actual CFD yielded efficiencies in Fig.~\ref{figure_new3}(b). The $R^2$ value of this fit is 0.9929, indicating that this response surface can be used for subsequent inference. 

We repeated the aforementioned numerical results on the data-sets associated with Blades B and C, achieving similar results. In Sec.~\ref{sec:four} we expand upon these findings.

%% file: sec-6.tex
\section{Supporting Multi-point Design via Dimension Reduction}
\label{sec:four}
Our prior analysis of Blade C was carried out on its sea-level condition, i.e.~the inlet stagnation pressure and the exit flow capacity boundary conditions were based on operation at sea-level. We apply the same variable projection workflow along the cruise working line, requiring another 200 CFD evaluations. It is important to note that the baseline geometries for both cruise and sea-level conditions are different, and we match the \emph{hot running} shape of the datum blade (no perturbations) at these conditions. Thus, in this section we are largely interested in a comparative study of the design space perturbations as opposed to absolute geometry descriptions. 

\begin{figure}[h]
\centering
\includegraphics[scale=0.4]{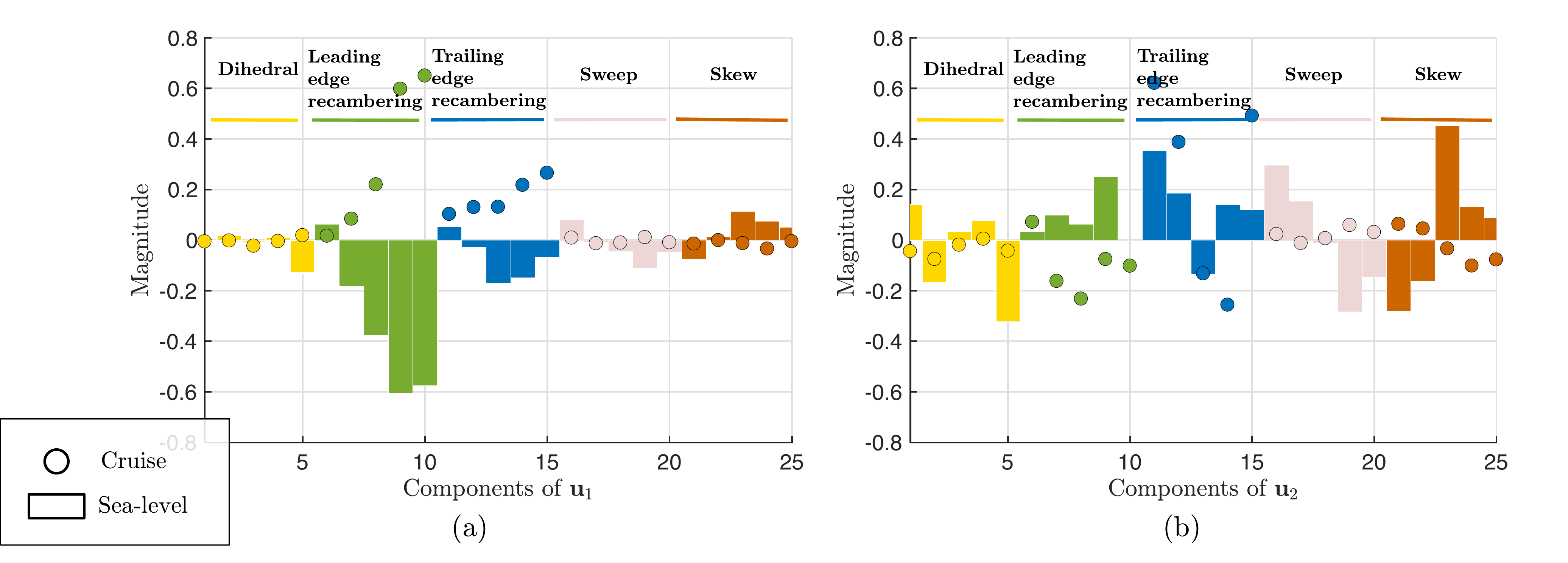}
\caption{Comparing the cruise and sea-level subspaces $\mU$: components of $\vu_{1}$ are shown in (a) and components of $\vu_{2}$ in (b).}
\label{compare_sub}
\end{figure}

A comparison of the components of the two subspaces is shown in Fig.~\ref{compare_sub}, where the markers in the figures denote the 25 design variables. Although the subspaces are different (varying more so along $\vu_{2}$ than along $\vu_{1}$), we can compare designs that have high efficiencies at sea-level with designs that have high efficiencies at cruise and vice-versa. 

\subsection{Generating constrained designs}
\noindent First, we define the matrix 
\begin{equation}
\mW=\left[\begin{array}{cc}
\mU & \mV\end{array}\right]
\end{equation}
where $\mV=\mathsf{null}\left(\mU^{T}\right)$, in which $\mathsf{null}$ represents the nullspace, and $\mU$ is the dimension reducing subspace obtained from variable projection. As $\mW$ is orthogonal, we can write 
\begin{align}
\begin{split}
& -\boldsymbol{1}\leq \vx \leq \boldsymbol{1} \\
\Rightarrow & -\boldsymbol{1}\leq  \mW \mW^{T} \vx \leq \boldsymbol{1} \\
\Rightarrow & -\boldsymbol{1}\leq  \mU \mU^{T} \vx  + \mV \mV^{T} \vx  \leq \boldsymbol{1} \\
\Rightarrow & -\boldsymbol{1}\leq  \mU \vy  + \mV \vz  \leq \boldsymbol{1}
\end{split}
\label{equ_chebyshev}
\end{align}
where $\vy = \mU^{T} \vx$ and $\vz = \mV^T \vx$. This implies that one can generate numerous $\vx$ that have the same coordinates $\vy$ but different coordinates $\vz$, under the constraint that $\vx$ still lies within the original bounding box. This can be conveniently expressed as a linear programming problem where for the same coordinate $\vy$ we wish to generate numerous $\vz$.

\subsection{Multi-point results for Blade C}
\begin{figure}
\centering
\includegraphics[scale=0.4]{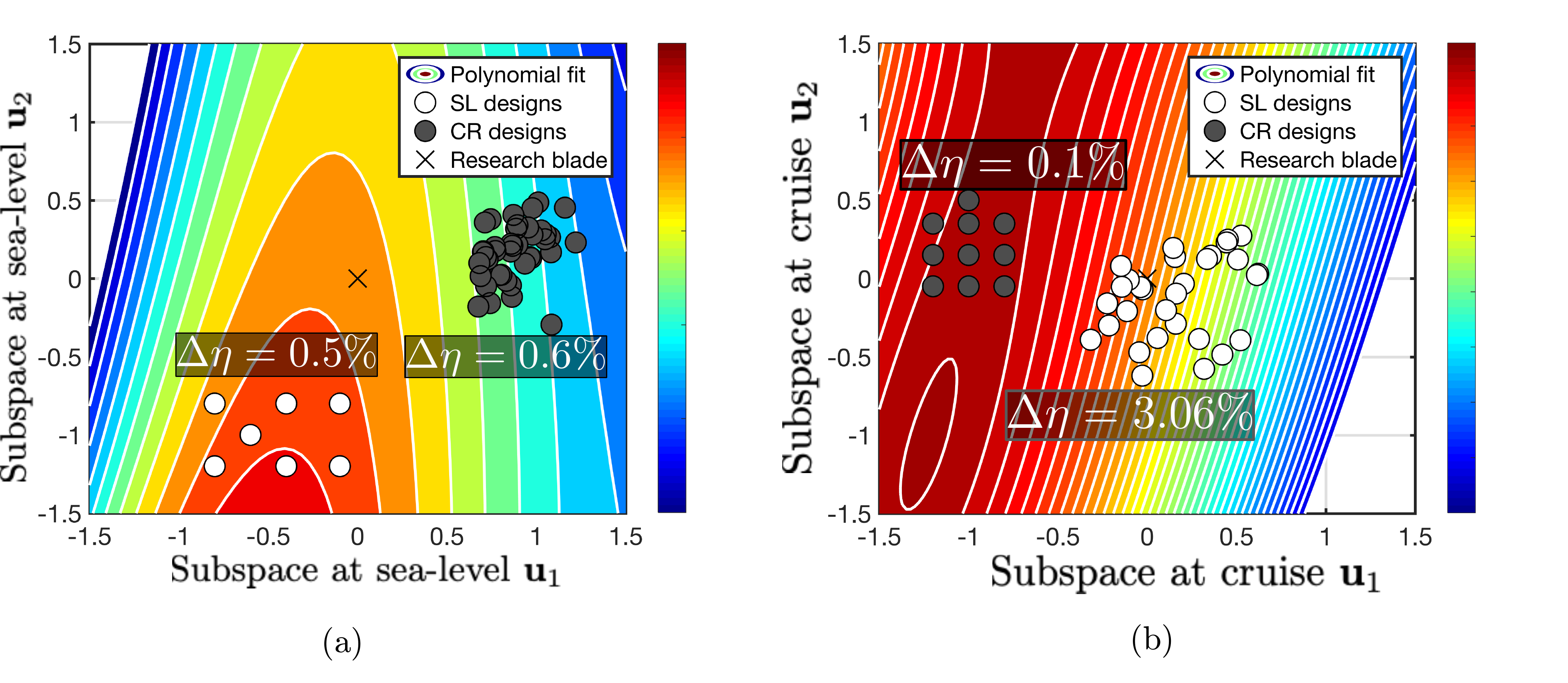}
\caption{Contours of efficiency as a function of its subspace $\mU$ for Blade C at two different conditions: (a) sea-level; (b) cruise. The difference between successive contour levels is 0.2$\%$}
\label{final_fig}
\end{figure}

We apply such a linear programming strategy to generate samples that map to the different subspaces; our results are summarized in Fig.~\ref{final_fig}. In Fig.~\ref{final_fig}(a) we plot the contours of sea-level efficiency on the sea-level subspace (essentially a different perspective on Fig.~\ref{figure_new3}(a), but for Blade C), with the initial geometry of Blade C denoted by the $\times$ marker and 7 sample 2D coordinates denoted by the white circular markers. For each marker we generate four 25D vectors that satisfy Eq.~\eqref{equ_chebyshev}, i.e.~they have different values along $\vz$ but the same value along $\vy$. These 28 new designs are then projected on the cruise efficiency subspace, shown in Fig.~\ref{final_fig}(b). We repeat this exercise for 10 high-efficiency designs in the cruise subspace, shown by the dark gray circular markers. It is clear that designs that attain higher efficiencies at cruise have a lower efficiency at sea-level and vice versa. Further, it is interesting to note that the original design reflects a healthy compromise between maximum efficiency at sea-level and at cruise. It is important to emphasize that we can infer the efficacy of various design configurations at these two operating points by visualizing the location of their $\vy$ coordinates along the two subspaces.

In Fig.~\ref{final_fig}, we show the range (maximum - minimum) in efficiency for sea-level and cruise designs on both subspaces. On the sea-level subspaces, both the cruise and sea-level designs selected have an efficiency range of 0.6$\%$ and 0.5$\%$ respectively. These same design perturbations, when mapped on to the cruise subspace yield efficiency ranges of 0.1$\%$ and 3.06$\%$ respectively. There are two important remarks to make here. The first (and the obvious one) is that design perturbations aimed at improving the efficiency at sea-level are likely to have a detrimental effect at cruise. Second, higher efficiency designs at cruise (see dark gray markers in Fig.~\ref{final_fig}(b)) warrant further deviations from the datum design; inferred from their coordinates along $\vu_{1}$. Analogously, higher efficiency designs at sea level (see white markers in Fig.~\ref{final_fig}(a)) have comparatively reduced design excursion. However, despite this, these designs yield a greater efficiency penalty. This leads to the conclusion that greater design perturbations can potentially yield more favorable designs---a particularly important point when investigating the impact of manufacturing variations in design.

\begin{figure}
\centering
\includegraphics[scale=0.5]{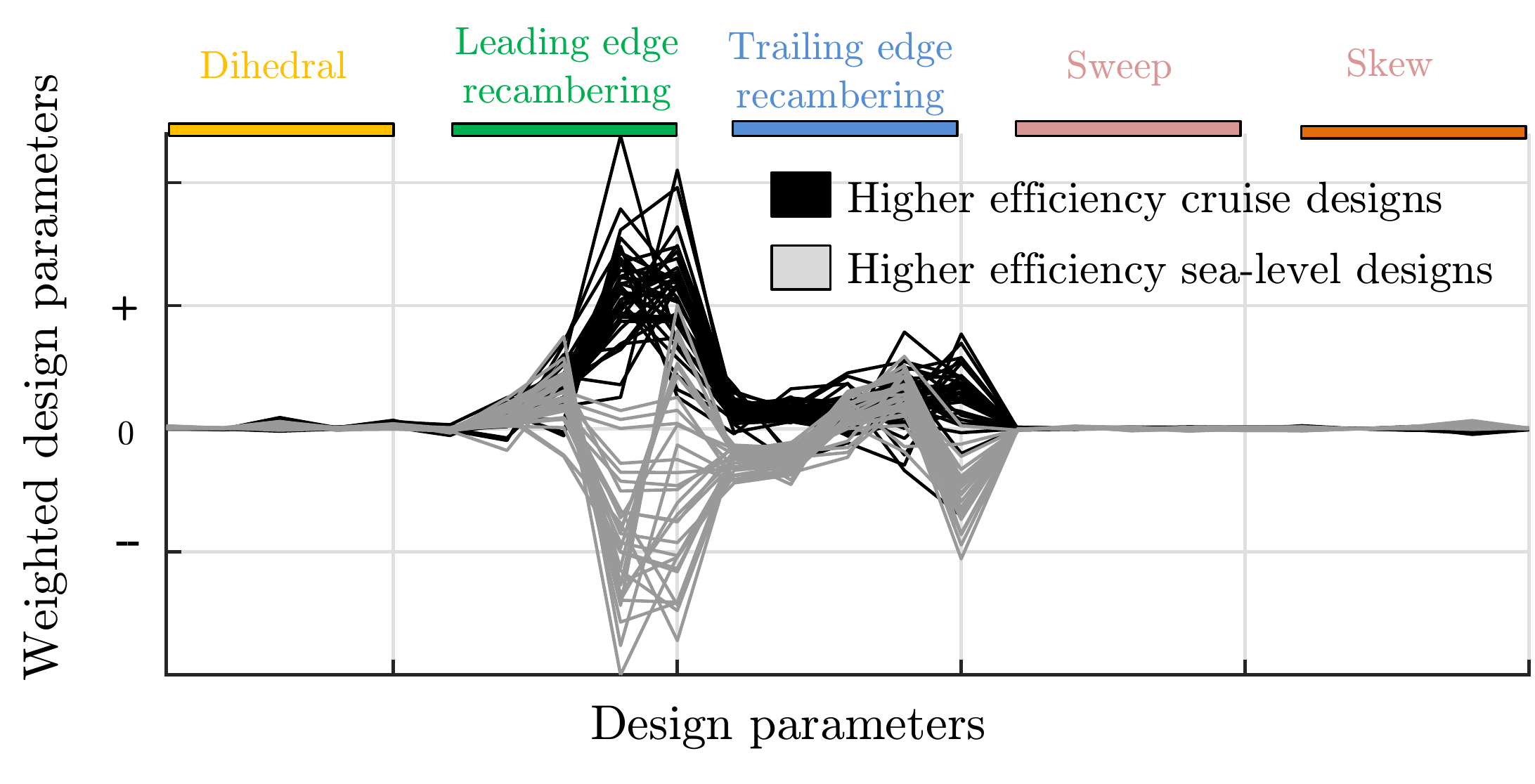}
\caption{A parallel coordinates representation for higher efficiency designs at both cruise and sea-level conditions.}
\label{final_F1}
\end{figure}

\begin{figure}
\centering
\includegraphics[scale=0.3]{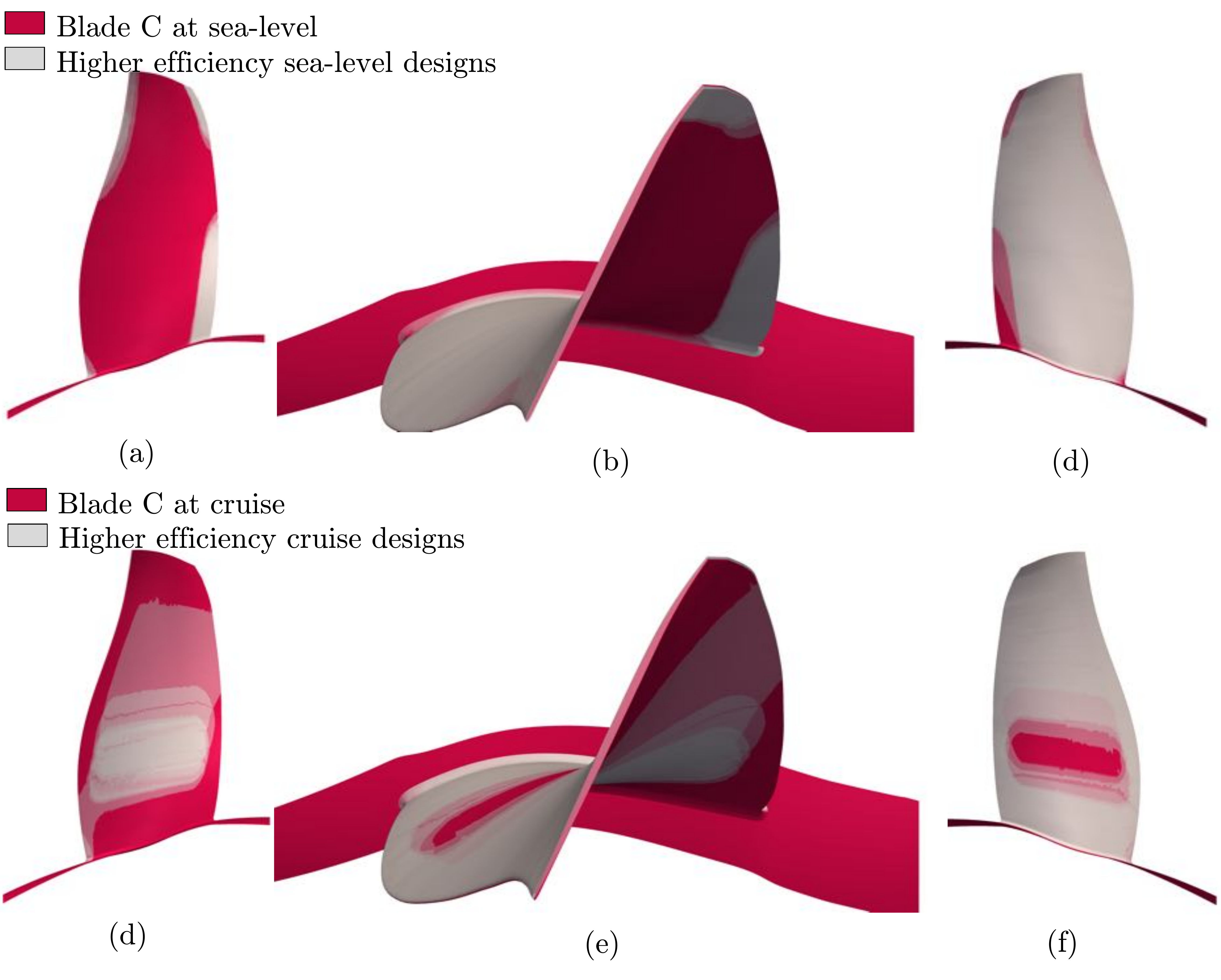}
\caption{Comparison of higher efficiency designs at sea-level in (a, b and c) and at cruise in (d, e and f). Higher efficiency designs are shown in gray with the baseline shown in magenta. Geometries are scaled for proprietary reasons.}
\label{final_F2}
\end{figure}

To gain a better understanding of the differences between the geometries, consider the parallel coordinates plot shown in Fig.~\ref{final_F1}. Here the vertical axis has been scaled by multiplying the dimensionalized design variables of the various blades with the absolute value of the cruise subspace $\vu_{1}$ shown previously in Fig.~\ref{compare_sub}(a). We remark here that this is purely to distinguish between the leading and trailing recambering dofs, which were found be to more important than the other three dofs. A distinct trend in parallel coordinates is readily apparent. Higher efficiency designs at cruise have both---on average---positive leading and trailing edge recambering. However, for higher efficiency designs at sea-level, the trend is reversed. The blade shapes for these cases are shown in Fig.~\ref{final_F2}. In addition to recambering, minor differences in dihedral are also evident. 

%% file: sec-7.tex
\section{Conclusions}
Our first goal in this paper was to study numerous dimension reduction strategies that can identify dimension reducing subspaces for 3D turbomachinery simulations. We analyzed four sufficient dimension reduction methods---namely SIR, SAVE, pHd and CR---and polynomial variable projection. In our numerical experiments we found the latter to parsimoniously identify a dimension reducing subspace---by requiring approximately 200 fewer CFD evaluations compared to a global quadratic model based active subspace technique. Finally, we demonstrated how designs that offer a healthy compromise between cruise and sea-level conditions can be easily found by visually inspecting their subspaces. This can empower further design trade-off studies and serve as an important tool for the blade designer. 

\section{Acknowledgments}
The authors are grateful to Rolls-Royce plc for permission to publish this work. The opinions and views expressed in this paper are those of the authors and not necessarily those of Rolls-Royce.